\PassOptionsToPackage{table,xcdraw}{xcolor}
\documentclass{article}
\usepackage{PRIMEarxiv}
\usepackage{fontawesome5}
\usepackage{url,lineno,microtype}
\usepackage{amsmath,amssymb,amsthm}
\usepackage{multirow}
\usepackage{tabularray}
\usepackage{array}
\usepackage{float}
\usepackage{multicol}
\usepackage{url}
\usepackage{xcolor}
\usepackage{tikz}
\usepackage{wrapfig}
\usepackage{pict2e}
\usepackage{textgreek}
\usepackage{datetime, scrdate}
\usepackage{tcolorbox}
\usepackage{comment}
\usepackage{adjustbox}
\usepackage{booktabs,xltabular}
\usepackage{tabularx,tabulary}
\usepackage{iftex}
\usepackage[square,sort,comma,numbers]{natbib}
\usepackage{stfloats}
\usepackage{microtype}
\usepackage{tablefootnote}
\usepackage{wrapfig}
\usepackage[colorlinks=true,linkcolor=black,anchorcolor=black,filecolor=black,menucolor=black,runcolor=black,urlcolor=black]{hyperref}
\usepackage[normalem]{ulem}

\useunder{\uline}{\ul}{}

\definecolor{light-yellow}{RGB}{255, 217, 102}
\definecolor{light-red}{RGB}{221, 126, 107}
\definecolor{light-blue}{RGB}{109, 158, 235}
\definecolor{our-turquoise}{RGB}{54, 255, 248}
\definecolor{our-green}{RGB}{103, 253, 154}
\newcommand{\tikzcircle}[2][red,fill=red]{\tikz[baseline=-0.5ex]\draw[#1,radius=#2] (0,0) circle ;}%
\newcommand*\circled[1]{\tikz[baseline=(char.base)]{
            \node[shape=circle,draw,inner sep=1pt] (char) {#1};}}

\def\BibTeX{{\rm B\kern-.05em{\sc i\kern-.025em b}\kern-.08em
    T\kern-.1667em\lower.7ex\hbox{E}\kern-.125emX}}

\title{A Matter of Annotation: An Empirical Study on In Situ and Self-Recall Activity Annotations from Wearable Sensors}

\author{
  Alexander Hoelzemann \\
  Ubiquitous Computing \\
  University of Siegen \\
  Siegen\\
  \texttt{alexander.hoelzemann@uni-siegen.de} \\
   \And
  Kristof Van Laerhoven \\
  Ubiquitous Computing \\
  University of Siegen \\
  Siegen\\
  \texttt{kvl@eti.uni-siegen.de} \\
}

\begin{document}
\maketitle

\begin{abstract}
Research into the detection of human activities from wearable sensors is a highly active field, benefiting numerous applications, from ambulatory monitoring of healthcare patients via fitness coaching to streamlining manual work processes.

We present an empirical study that evaluates and contrasts four commonly employed annotation methods in user studies focused on in-the-wild data collection. For both the user-driven, in situ annotations, where participants annotate their activities during the actual recording process, and the recall methods, where participants retrospectively annotate their data at the end of each day, the participants had the flexibility to select their own set of activity classes and corresponding labels.

Our study illustrates that different labeling methodologies directly impact the annotations' quality, as well as the capabilities of a deep learning classifier trained with the data.
We noticed that in situ methods produce less but more precise labels than recall methods. Furthermore, we combined an activity diary with a visualization tool that enables the participant to inspect and label their activity data. Due to the introduction of such a tool were able to decrease missing annotations and increase the annotation consistency, and therefore the F1-Score of the deep learning model by up to 8\% (ranging between 82.1 and 90.4 \% F1-Score).
Furthermore, we discuss the advantages and disadvantages of the methods compared in our study, the biases they could introduce, and the consequences of their usage on human activity recognition studies as well as possible solutions.
\end{abstract}

\keywords{Human-centered computing \and ubiquitous computing \and user-driven study \and annotations}

section{Introduction}

Sensor-based activity recognition is one of the research fields of Pervasive Computing developed with enormous speed and success by industry and science and influencing medicine, sports, industry, and therefore the daily lives of many people.
However, current available smart devices are mostly capable of detecting periodic activities like simple locomotions. In order to recognize more complex activities a multimodal sensor input, such as \citep{roggen2010collecting}, and more complex recognition models are needed.
Many of the published datasets are made in controlled laboratory environments. Such data does not have the same characteristics and patterns as data recorded in-the-wild. Data that belongs to similar classes but is recorded in an uncontrolled versus controlled environment can differ significantly since it contains more contextual information \citep{mekruksavanich2021recognition} . Furthermore, study participants tend to control their movements more while being monitored \citep{friesen2020all}.
The recording of long-term and real-world data is a tedious, time-consuming, and therefore a non-trivial task. Researchers have various motivations to record such datasets but the technical hurdles are still high and problems during the annotation process occur regularly.

In Human Activity Recognition research, capturing long-term datasets presents a challenge: balancing precise labeling with minimal participant burden. Sole reliance on self-recall methods, like activity diaries e.g., \citep{zhao2013healthy}, often leads to imprecise time indications that may not accurately reflect actual activity duration. Such incorrectly or noisy labeled data later on leads to a trained model that is less capable of detecting activities reliably \citep{natarajan2013learning}, due to unwanted temporal dependencies learned by wrongly annotated patterns \citep{bock2022investigating}.

The field of HAR is witnessing a growing emphasis on real-world and long-term activity recognition. This focus stems from the need to address current limitations and achieve reliable recognition of complex daily activities. Existing long-term datasets often rely heavily on self-recall methods or additional tracking apps. These apps can set labels either automatically \citep{akbari2021facilitating} or require manual selection \citep{cleland2014evaluation}. However, such approaches present challenges, leading many researchers to favor controlled environments for data collection. As a consequence, the number of publicly available "in-the-wild" datasets remains limited.
\\
\textbf{Contribution:} Our study focuses on the evaluation of 4 different annotation methods for labeling data in-the-wild: \circled{1} In situ \textit{(lat. on site or in position)}  with a button on a smartwatch, \circled{2} in situ with the app Strava \footnote{\url{https://www.strava.com/}} (an app that is available for iOS and Android smartphones), \circled{3} pure self-recall (writing an activity diary at the end of the day), and \circled{4} \textit{time-series} assisted self-recall with the MAD-GUI \citep{ollenschlager2022mad}, which displays the sensor data visually and allows to annotate it interactively. Our study was conducted with 11 participants, 10 males, and 1 female, over 2 weeks. Participants wore a Bangle.js Version 1\footnote{\url{https://www.espruino.com/banglejs}} smartwatch on their preferred hand, used Strava, and completed self-recall annotations every evening.
In the first week, the participants were asked to write an activity diary at the end of the day without any helping material and additionally using two user-initiated methods (\textit{in situ button} and \textit{in situ app}) to manually set labels at the start and beginning of each activity. In the second week, the participants were given an additional visualization of the sensor data with an adapted version of the MAD-GUI annotation tool. With the help of this, participants then were instructed to label their data in hindsight with the activity diary as a mnemonic aid. Given labels from both weeks were compared to each other regarding the quality through visual inspection and statistical analysis with regard to the consistency and quantity of missing annotations across labeling methods.

The participants in this study were given the freedom to self-report their activity classes based on the diverse range of pursuits encompassed within their daily lives. Consequently, the resulting dataset exhibits a heterogeneous composition, comprising both commonplace, routine activities such as \textit{walking, driving}, and \textit{eating}, as well as more specialized and niche activities like \textit{badminton, yoga, horse\_riding}, and {gardening}.

Furthermore, we used a Shallow-DeepConv(LSTM) architecture, see \citep{bock2021improving} and \citep{ordonez_2016}, and trained models with a Leave-One-Day-Out cross-validation method of 6 previously selected subjects and each annotation method.\\
\textbf{Impact:} Annotating data, especially in real-world environments, is still very difficult and tedious. Labeling such data is always a trade-off between accuracy and workload for the study participants or annotators. We raise awareness among researchers to put more effort into exploring new annotation methods to overcome this issue. Our study shows that different labeling methodologies have a direct impact on the quality of annotations. With the deep learning analysis, we prove that this impacts the model capabilities directly. Therefore, we consider the evaluation of frequently used annotation methods for real-world and long-term studies to be crucial to give decision-makers of future studies a better base on which they can choose the annotation methodology for their study in a targeted way.

\section{Related work}
A very limited number of datasets are currently publicly available which were recorded in the wild, e.g. \citep{berlin2012detecting,vaizman2018extrasensory, thomaz2015practical,sztylerOnBodyLocalizationWearable2016,gjoreski2018university}. \citep{gjoreski2018university} and \citep{sztylerOnBodyLocalizationWearable2016} were captured in naturalistic settings, but the participants were equipped with multiple sensors on various body locations and were filmed by a third party during the exercises. Such visible equipment and the presence of an observer could potentially introduce a behavior bias \citep{yordanova2018ardous} in the data, as it may alter participants' movement patterns due to the constant reminder that they are participating in a study \citep{friesen2020all}. Furthermore, multimodal datasets recorded with multiple body-worn sensors, rather than a single Inertial Measurement Unit (IMU), have faced the challenge of proper inter-sensor synchronization \citep{hoelzemann2019using}.
A comprehensive dataset encompassing a diverse range of classes, accurate annotations, and recorded by a single device that is nearly unnoticeable to the participant (and therefore unlikely to influence their behavior or movement patterns), is not yet publicly available due to the aforementioned obstacles.

According to \citep{stikic2011weakly} and later \citep{cleland2014evaluation}, we distinguish between 6 or 7, respectively, different methods and 2 environments (online/offline) of labeling data, the methods are (1) Indirect Observation, (2) Self-Recall, (3) Experience Sampling, (4) Video/Audio Recordings, (5) Time Diary, (6) Human Observer, (7) Prompted Labeling. \citep{cruz2019semi} uses 4 different categories to classify data labeling approaches, these are \textbf{(1) temporal (when)} - is the label conducted during or after the activity, \textbf{(2) annotator (who)} - is the label given by the individual itself or by an observer, \textbf{(3) scenario (where)} - is the activity labeled in a controlled (e.g laboratory) or uncontrolled (in-the-wild) environment, and \textbf{(4) annotation mechanism (how)} - is the activity labeled manually, semi-automatically or fully-automatically.
All labeling methods have their own benefits, and costs and come with a trade-off between required time and label accuracy.
However, not every method is suitable for long-term and in-the-wild recording data. \citep{reining2020annotation}, evaluated the annotation performance between 6 different human annotators of a MoCap (Motion Capturing) and IMU HAR Dataset for industrial deployment. They came to the conclusion that annotations were moderately consistent when subjects labeled the data for the first time. However, annotation quality improved after a revision by a domain expert.
In the following, we would like to go into more detail on what we consider to be the most important labeling methods for the specific field of activity recognition.

\subsection{Annotation Methods in Activity Recognition}
\textbf{Self-Recall} methodologies are generally called methods in which study participants have to remember an event in the past. This methodology is used, for instance, in the medical field (e.g. in the diagnosis of injuries \citep{valuri2005validity}), but also frequently in studies in the field of long-term activity recognition.
\citep{van2008using} used this method during a study in which participants were asked to label their personal daily data at the end of the day.
They noticed that the label quality depends heavily on the participant's recall and can therefore be very coarse.
During a study conducted by \citep{tapia2004activity}, every 15 minutes a questionnaire was triggered in which participants needed the answer multiple choice questions about which of 35 predefined activities were recently performed.

\textbf{App Assisted Labeling:} \citep{cleland2014evaluation} presented in 2014 the so called Prompted labeling. An approach that is already used by commercial smartwatches like the Apple Watch\footnote{\url{https://www.apple.com/watch/}}.
In this study user's were asked  to set a label for a time period which has been detected as an activity right after the activity stops.  \citep{akbari2021facilitating} leverages freely available Bluetooth Low Energy (BLE) information broadcasted by other nearby devices and combines this with wearable sensor data in order to detect context and direction changes. The participant is asked to set a new label whenever a change in the signal is detected. \citep{gjoreski2017versatile} published in 2017 the SHL dataset which contains versatile labeled multimodal sensor data that has been labeled using an Android application that asked the user to set a label whenever they detected a position change via GPS.
\citep{tonkin2018talk} presented a smartphone app that was used in their experimental smart home environment with which study participants were able to either use voice-based labeling, select a label from a list of activities ordered by the corresponding location or scan NFC tags that were installed at locations in the smart house. Similar to \citep{tonkin2018talk}, \citep{vaizman2018extrasensory} developed an open-source mobile app for recording sensor measurements in combination with a self-reported behavioral context (e.g. driving, eating, in class, showering). 60 subjects participated in their study. The study found that most of the participants preferred to fill out their past behavior through a daily journal. Only some people preferred to set a label for an activity that they are about to do. \citep{schroder2016tool} developed a web-based GUI which can either be used on a smartphone, tablet, or a PC to label data recorded in a smart home environment. However, it is important to mention that, According to \citep{cleland2014evaluation}, the process of continually labeling data becomes laborious for participants and can result in a feeling of discomfort.

\textbf{Unsupervised Labeling} is a methodology that uses clustering algorithms to first categorize new samples without deciding yet to which class a sample belongs. \citep{leonardis2002multiple} presented in 2002 the concept of finding multiple subsets of eigenspaces where, according to \citep{huynh2008human}, each of them corresponds to an individual activity. Huynh uses this knowledge to develop the eigenspace growing algorithm, whereby, \textit{growing} refers to an increasing set of samples as well as to increasing the so-called \textit{effective dimension} of a corresponding eigenspace. Based on the reconstruction error (when a new sample is projected to an eigenspace), the algorithm tries to find the best-fitting representation of a sample with minimal redundancy.
\citep{hassan2021autoact} recently published a methodology that uses the Pearson Correlation Coefficient to map very specific labels of a variety of datasets to 4 meta labels (inactive, active, walking, and driving) of the ExtraSensory Dataset \citep{vaizman2018extrasensory}.\\
\textbf{Human-in-the-Loop (Labeling)} is a collective term for methodologies that integrates human knowledge into their learning or labeling process.
Besides of being applied in HAR research, such techniques are often used in Natural Language Processing (NLP) and according to \citep{wu2022survey} the NLP community distinguishes between entity extraction \citep{gentile2019explore, zhang2019invest}, entity linking \citep{klie2020zero}, Q\&A tasks \citep{wallace2019trick} and reading comprehension tasks \citep{bartolo2020beat}.

\textbf{Active Learning} is a machine learning strategy that currently receives a lot of attention in the HAR community. Such strategies involves a Human-in-the-Loop for labeling purposes. In the first step the learning algorithm automatically identifies relevant samples of a dataset which are posteriorly queued to be annotated by an expert. Incorporating a human guarantees high quality labels which directly leads to a better performing classifier.
Whether a sample is determined to be relevant, and as well the decision to whom it may get presented for annotation purposes are the main focus of research in this field.  \citep{bota2019semi} presents a technique that relies on specific criteria defined by 3 different uncertainty-based selection functions to select samples that will be presented to an expert for labeling and then be propagated throughout the most similar samples. \citep{adaimi2019leveraging} benchmarks the performance of different Active Learning strategies and compared them, with regard to 4 different datasets with a fully-supervised approach. The authors came to the conclusion that Active Learning needs only 8\% to 12\% of the data to reach similar or even better results than a fully-supervised trained model. These results suggest that presenting pre-selected samples to a human for labeling purposes can reduce the amount of data needed to train a machine learning classifier significantly due to the increased quality of the labels.
\citep{miu2015bootstrapping} presented a system which used the Online Active Learning approach published by Scully \citep{sculley2007online} to bootstrap \citep{abney2002bootstrapping} a machine learning classifier. The publication presented a smartphone app that asked the user right after finishing an activity, which activity has been performed. Afterwards a small subset of the labeled data was used to bootstrap a personalized machine learning classifier.

\section{Methodology}
Our study is conducted with 11 participants, from which 10 are male and 1 is female. The participants are between 25 and 45 years old. Out of 11 participants, 6 are researchers in the field of signal processing and are used to read and work with sensor data. Participants were selected among acquaintances and colleagues.
\subsection{Study Setup}
\begin{figure}[!b]
    \centering
    \includegraphics[width=.9\columnwidth]{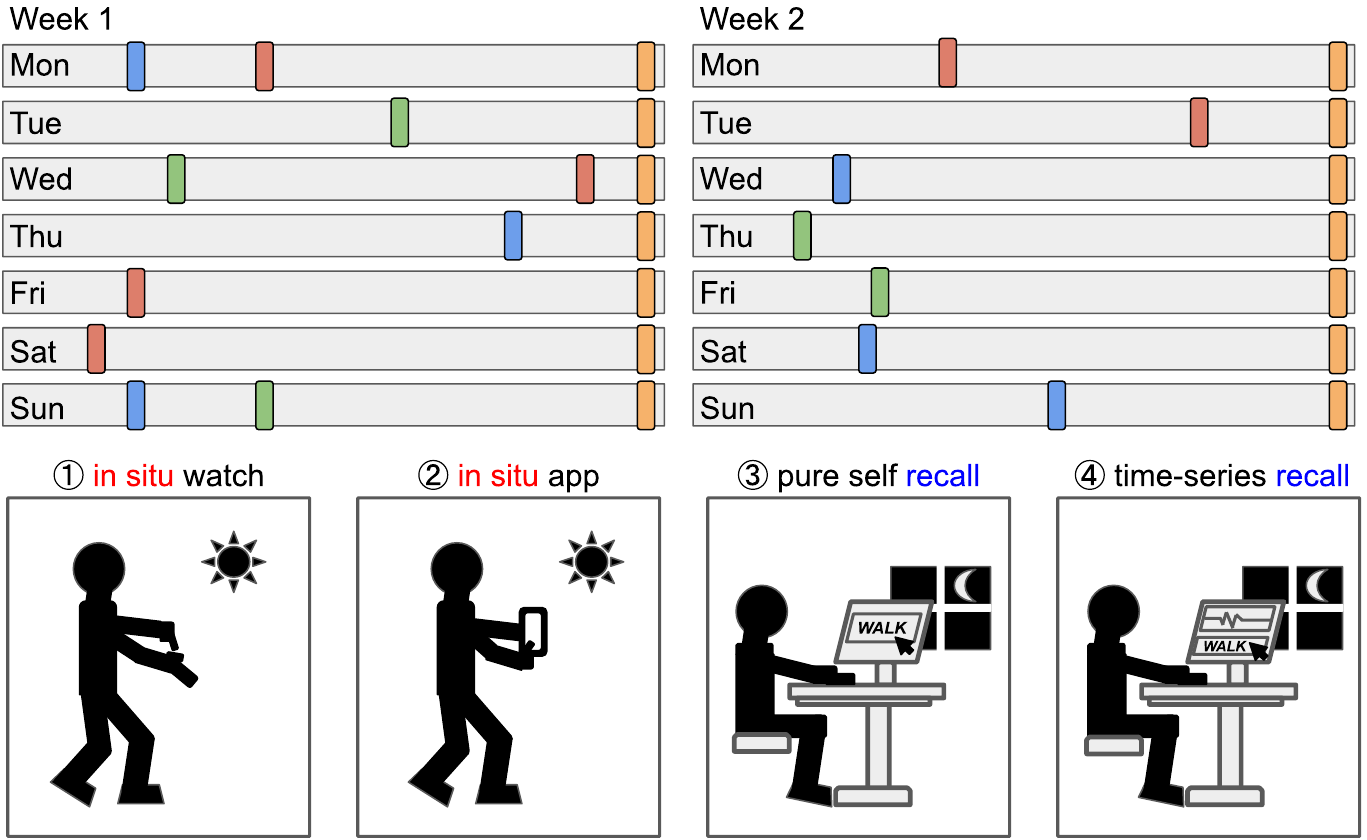}
    \caption[Comparing Annotation Methods: Overview of the 4 different annotation methods used in our study.]{The study participants collected data for 14 days in total and annotated the data with 4 different methods: Labeling \circled{1} in situ with a mechanical button, \circled{2} in situ with an app, \circled{3} by writing a pure self-recall diary and \circled{4} writing a self-recall diary assisted by visualization of their time-series data.}
    \label{fig:sketch}
\end{figure}

The study was conducted over a period of 2 weeks, during which participants wore an open-source smartwatch on their chosen wrist. Throughout the two-week study period, the participants were instructed to use 4 different labeling methods in parallel, as illustrated in Figure \ref{fig:sketch}. In the first week they were asked to use the \circled{1} \textit{in situ button}, \circled{2} \textit{in situ app}, and \circled{3} \textit{pure self-recall} methods. At the beginning of the 2nd week, we expanded the number of annotation methods with the \circled{4} \textit{time-series recall}. This annotation method combines the activity diary with a graphical visualization of the participants' daily data.

\circled{1} The Bangle.js smartwatch has 3 mechanical buttons on the right side of the case. These buttons are programmed to record the number of consecutive button presses per minute. The button-press annotation method captures the total number of button presses along with their corresponding timestamps, enabling the delineation of the beginning and end of an activity within the time-series data. However, this approach does not inherently assign a specific label or description to the identified activity segment. To address this limitation, we employed an inference strategy that leveraged the temporal alignment of the button-press data with the annotations obtained from other methods. By identifying segments with similar timestamps across multiple annotation modalities, we could infer the appropriate label for the button-press annotations.

\circled{2} In addition, the participants were asked to track their activities with the smartphone app Strava. Strava is an activity tracker that is available for Android and iOS and freely downloadable from the app stores. The user can choose from a variety of predefined labels and start recording. Recording an activity starts a timer that runs until the user stops it. The time as well as the GPS position of the user during the activity is tracked and saved locally.

\circled{3} The \textit{pure self-recall} methods consist of writing an activity diary on a daily basis at the end of the day. The participants were explicitly told that they should only write down the activities that they still remember 2 hours after the measurement stopped.

\circled{4} The \textit{time-series recall} method can be seen as a combination of an activity diary and a graphical representation of the raw sensor data. For visualization and labeling purposes, we provided the participants with an adapted version of the MAD-GUI. The GUI was published by \citep{ollenschlager2022mad} in 2022 and is a generic open-source Python package. Therefore, it can be integrated into one's project. Our adaptions to the package are available for download from a GitHub repository\footnote{\url{https://github.com/ahoelzemann/mad-gui-adaptions/}}. It contains changes to the data loader, the definition of available labels, and color settings for displaying the 3D raw data.

\textbf{Annotation Guidelines:} The participants were provided with guidelines that instructed them to document recurring daily activities, encompassing both sports and activities of daily living, that exceeded a duration threshold of 10 minutes. However, the annotation process was deferred until approximately 2 hours after the cessation of the recording session. This temporal offset was implemented to allow for a reasonable time buffer, enabling participants to consolidate their experiences. For instance, if a daily recording concluded at 7 pm, the participant would typically annotate their data around 9 pm, allowing for a 2-hour interim period.
Each of these annotation methods represents a layer of annotation that is used for the visual, statistical, and deep learning evaluation. Figure \ref{fig:sketch} illustrates the overall concept.

\textbf{Annotation Process:} To capture realistic daily data reflecting participants' natural routines, we granted them complete autonomy in choosing their activity classes. Participants were not restricted to a specific activity protocol; instead, we left the decision of what to label entirely to their judgment. During the study's first week, participants employed methods \circled{1} - \circled{3} concurrently. In the second week, method \circled{4} was introduced for them to utilize alongside the existing methods. The labels provided by the participants were later interpreted by the researchers and, when necessary, categorized into meta-classes. However, whenever a participant was specific about the activity performed, their label was not summed up into a meta-class. For example, activities such as \textit{yoga, badminton}, or \textit{horse\_riding} were not combined under the meta-class \textit{sport}.

\subsection{Hardware}
Participants wore the commercial open-source smartwatch Bangle.js Version 1 with our open-source firmware\footnote{Our smartwatch firmware is made publicly available at: \url{https://github.com/kristofvl/BangleApps/tree/master/apps/activate}} installed.
The device comes with a Nordic 64MHz nRF52832 ARM Cortex-M4 processor with Bluetooth LE, 64kB RAM, 512kB on-chip flash, 4MB external flash, a heart rate monitor, a 3D accelerometer, and a 3D magnetometer.
Our firmware only uses the 3D accelerometer and provides the user with the basic functions of a smartwatch, like displaying the time and counting steps. The data is recorded with 25Hz, a sensitivity of ±8g and saved on the devices' memory with a delta compression algorithm. Therefore, we are able to save up to 8-9 hours (depending on how much of the data could be compressed) of data with the given parameters. The smartwatch stops recording as soon as the memory is full. At the end of the day, the participants need to upload their daily data and program the starting time for the next day using our upload web-tool\footnote{Our web-tool is made publicly available at:
\url{https://ubi29.informatik.uni-siegen.de/upload/}}.

\section{Statistical Analysis}
The labels were statistically analyzed based on their consistency using the Cohen $\kappa$ score as well as the number of missing annotations across all methods. The Cohen $\kappa$ score describes the agreement between two annotation methods, which is defined as follows $\kappa = (p_0 - p_e)/(1 -p_e)$, see \citep{artstein2008inter} and \citep{reining2020annotation}. Where $p_0$ is the observed agreement ratio and $p_e$ is the expected agreement if both annotators assign labels randomly. The score shows how uniform two different annotators labeled the same data. For calculation purposes, an implementation provided by Scikit-Learn \citep{scikit-learn-cohen}, was used.
Furthermore, missing annotations across methods are measured as the percentage of missing or incomplete annotations. The annotations of all methods were first compared with each other and matched based on the given time indications.  Annotations that could not be assigned or were missing were marked accordingly and are the base for calculating this indicator, see Section \ref{sec:missing_annotation} for more information.

We used a similar representation as \citep{brenner1999errors} to visualize the matches among labeling methods. In this study, the authors compared genome annotations labeled by different annotators with regard to their error scores between different annotators.
\section{Effects on Deep Learning Performance}
\begin{figure}[!t]
    \centering
    \includegraphics[width=\textwidth]{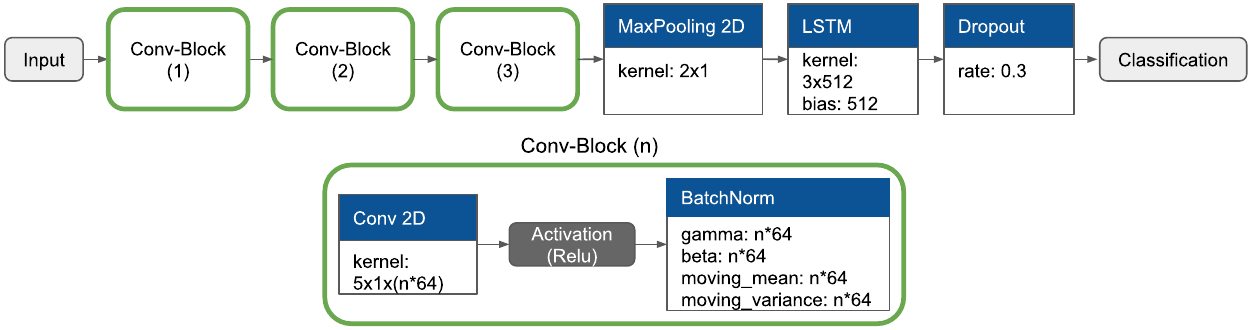}
    \caption[Comparing Annotation Methods: Used DeepConvLSTM architecture.]{The architecture consists of an \textbf{Input} Layer with the kernel-size 10 (window\_size) x 10 (filter\_length) x 3 (channels). The data is passed into 3 concatenated \textbf{convolutional blocks}, followed by a \textbf{MaxPooling} (kernel 2x1) where 50\% of the data is filtered.
    The convolutional block consists of a convolutional layer with a variable kernel size of 5x1x(n*64) following a ReLU activation function and a BatchNorm-Layer. We decided to use a single LSTM-Layer with the size of 512 units, as mentioned by \citep{bock2021improving}, which is followed by a Dropout-Layer that filters 30\% of randomly selected samples of the window.}
    \label{fig:architecture}
\end{figure}

The deep learning analyses are performed using the DeepConvLSTM architecture \citep{ordonez_2016} which is based on a Keras implementation of \citep{hoelzemann2020digging}. We did not perform hyperparameter tuning because it would involve a considerable amount of additional workload, since we trained 64 models independently during the evaluation. We therefore decided to opt out of the architecture with regards to efficiency rather than optimal classification results. Additionally, we don't expect that the actual experiment - evaluating different annotation methods - would benefit from hyperparameter tuning or gain any significant information and insights. Instead, we use the default hyperparameters provided by the authors. These are depicted in the Figure \ref{fig:architecture}.
Furthermore, we reduce the number of LSTM layers to one and instead increase the number of hidden units of the only LSTM layer to 512. According to \citep{bock2021improving}, this modification decreases the runtime up to 48 \% compared to a two-layered DeepConvLSTM while significantly increasing the overall classification performance on 4/5 publicly available datasets: \citep{reyes2016transition}, \citep{sztylerOnBodyLocalizationWearable2016}, \citep{roggen2010collecting}, \citep{scholl2015wearables}, \citep{stisen2015smart}. LSTM-Layers in general are important if the dataset contains sporadic activities \citep{bock2022investigating}. However, our dataset does not and our evaluation aims to identify long periods of periodic activities, like walking or running. For this reason, we can conclude that additional LSTM layers are not needed.
The implementation of \citep{hoelzemann2020digging} incorporates BatchNormalization layers after each Convolutional layer, as well as MaxPooling for the transition between the final convolutional block and the LSTM layer, and a Dropout layer before classification. Each Convolutional layer employs a ReLU activation function.
The inclusion of the BatchNormalization layers serves to accelerate training and mitigate the detrimental effects of internal covariate shift, as discussed further in \citep{ioffe2015batch}.

\textbf{Preprocessing: }To prepare the data for neural network training, we perform two preprocessing steps. First, we address minor inconsistencies in the device's sampling rate. The original data was collected at a rate of 12.5 Hz. However, for optimal performance with neural networks, a consistent and regular sampling rate is preferred. To achieve this, we upsample the data by a factor of two, resulting in a constant frequency of 25 Hz. This upsampling process essentially inserts additional data points between the existing ones, effectively increasing the resolution of the signal. The second preprocessing step involves rescaling the accelerometer data to a range between -1 and 1.

\begin{figure}[H]
\centering
    \includegraphics[width=1.0\columnwidth]{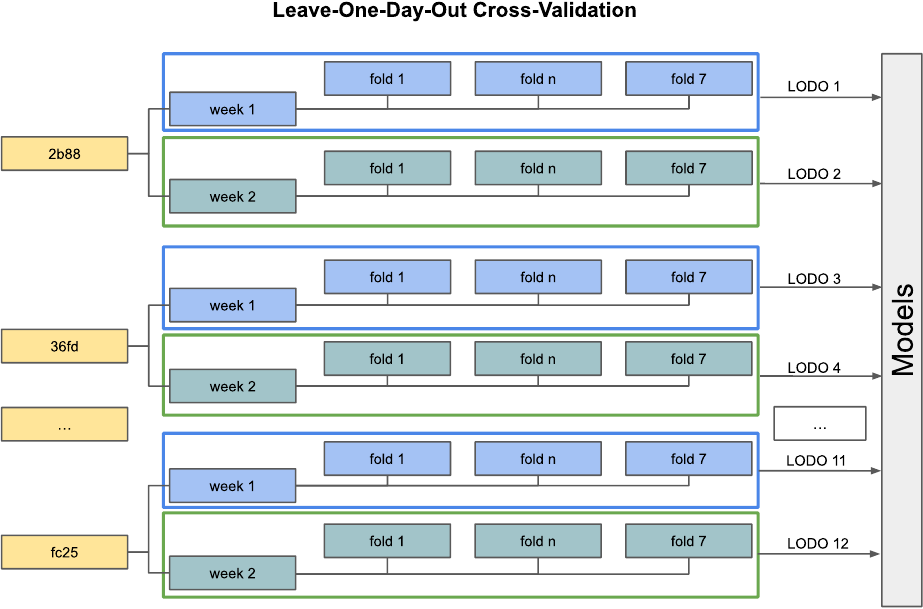}
    \caption[Comparing Annotation Methods: Leave-One-Day-Out Cross Validation.]{Leave-One-Day-Out Cross Validation. The models are personally trained for every participant and are not intended to generalize across all study participants. Instead, a generalization across all days of one week is desired.}
    \label{fig:lodo}
\end{figure}

\textbf{Leave-One-Day-Out Cross-Validation:} Figure \ref{fig:lodo} illustrates the train and test setting for the deep learning model. Instead of following the traditional Leave-One-Subject-Out strategy, we adapted it to our needs by using one day of the week for testing and training on the remaining days for each study participant and week.
This approach was necessitated by the unique characteristics of our dataset. It consists of a predominant \textit{void} class and a small number of samples per activity class and participant. To mitigate the issue of an disproportionately large \textit{void} class, we trained our model with balanced class weights.
By not limiting the participants in their choice of daily activities and not specifying predefined activity labels, we ended up with very unique sets of activities for each study participant. Given these circumstances, it is unrealistic to expect a model capable of generalizing across participants and days.
The Leave-One-Day-Out strategy aims to maintain the consistency of class labels within each day's data, providing a more cohesive and reliable dataset for training and evaluation purposes. This strategy also mitigates the potential impact of participant-specific biases or variations in class labeling, leading to a more robust and accurate model.
Furthermore, due to the in-the-wild recording setup, the intra-class differences \citep{bulling_2014} for comparatively simple activities, such as \textit{walking} or \textit{running} can be significant. Consequently, the impact of different labeling methods is expected to be more pronounced and visible in a personalized model compared to a generalized model.

\textbf{Post-Processing \& Classification:}
In the classification task, we initially segmented the data into fixed-length sliding windows of 2 seconds (50 samples). However, our objective extended beyond instantaneous classifications; we aimed to identify longer periods of recurring activities. To achieve this, we employed a post-processing technique involving a jumping window approach with a duration of 5 minutes. Within each 5-minute window, a majority vote was applied to the individual 2-second window predictions. The activity class with the highest number of occurrences within the 5-minute window was then assigned as the predominant activity for the entire window. This approach enabled us to capture sustained patterns of activities over extended periods, aligning with our goal of analyzing longer-term behavioral trends.
\section{Results}
Our participants were asked to annotate daily activities lasting more than 10 minutes. We did not limit them to a predefined set of classes; they independently decided on labels for their activities. After normalizing the labels (e.g., changing "going for a walk" to "walking"), the participants assigned 26 different labels: \textit{laying, sitting, walking, running, cycling, bus\_driving, car\_driving, cleaning, vacuum\_cleaning, laundry, cooking, eating, shopping, showering, yoga, sport, playing\_games, desk\_work, guitar\_playing, gardening, table\_tennis, badminton, horse\_riding, cleaning, reading, weightlifting, manual\_work, dish\_washing}. Any unlabeled samples were classified as \textit{void}. However, after excluding infrequent or non-standalone classes (e.g., \textit{shopping} likely combines \textit{walking, standing, and sitting}), we reduced the dataset to 23 labels (22 activities plus \textit{void}): \textit{laying, sitting, walking, running, cycling, bus\_driving, car\_driving, vacuum\_cleaning, laundry, cooking, eating, shopping, showering, yoga, sport, playing\_games, desk\_work, guitar\_playing, gardening, table\_tennis, badminton, horse\_riding}. Nevertheless, the graphical representation of the distribution and the table in Section \ref{sec:class_distribution} include the full scope of classes.
\subsection{Class Distribution}\label{sec:class_distribution}

The class distribution reflects a broad range of activity classes that represent the daily lives of our participants.
\begin{figure}[!htb]
    \centering
    \includegraphics[width=\textwidth]{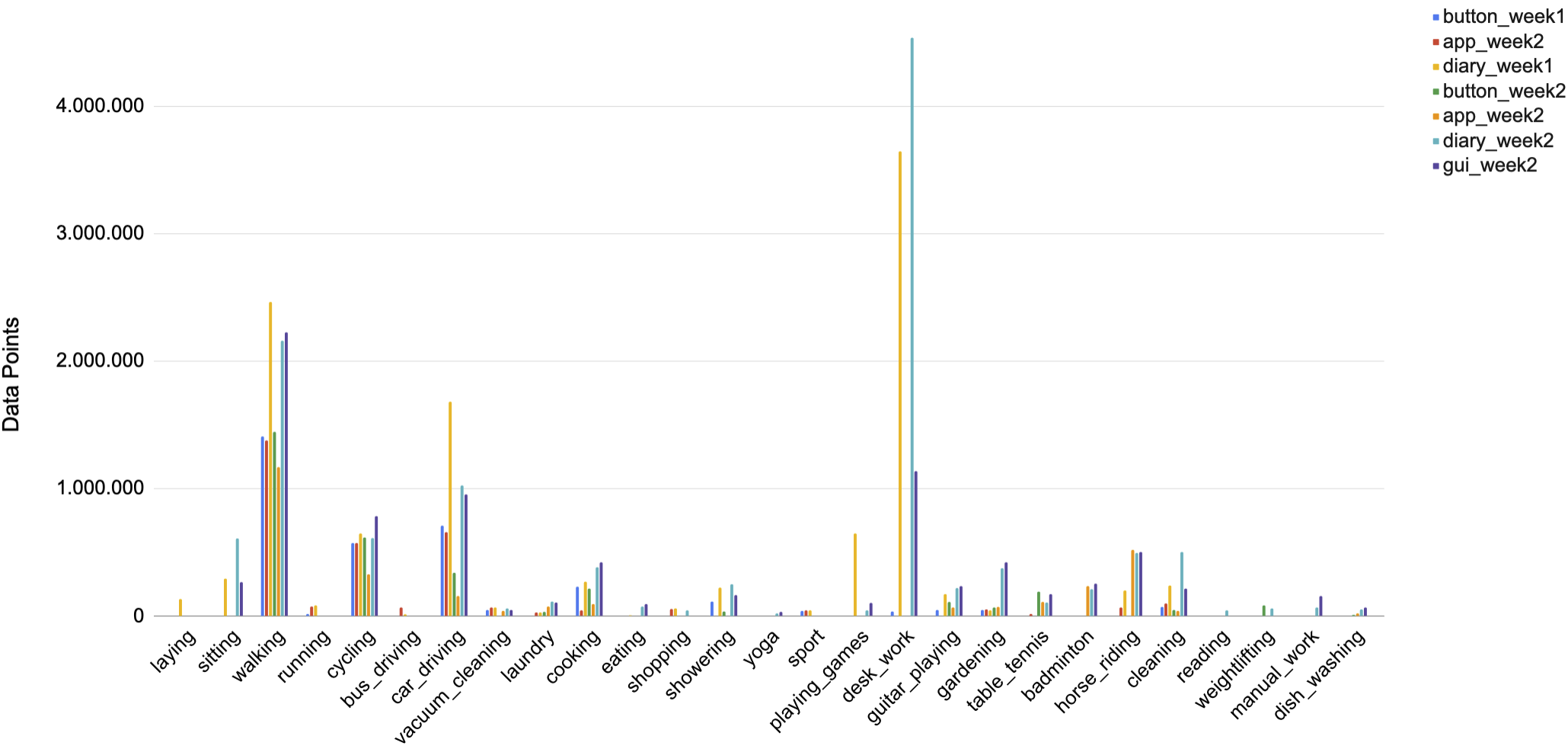}
    \caption[asdf.]{This figure illustrates the relative prevalence of various activity classes within the dataset, excluding the \textit{void} class, see Table \ref{tab:class_distribution} for details. The class labeled as \textit{void} represents the predominant category within the dataset, surpassing the frequency of the second most prevalent class, \textit{desk\_work}, by a substantial factor of 13. The figure illustrates a pronounced imbalance in the data distribution, both in terms of the annotation methodology employed and the distinct week-specific patterns observed in the annotation process. }
    \label{fig:class_distribution_lin}
\end{figure}
These classes remain primarily participant-specific due to the absence of a predefined annotation protocol, which allows participants the freedom to label activities according to their own interpretations.
Consequently, we decided against employing a Leave-One-Subject-Out evaluation method, as it might introduce inconsistencies in the dataset due to the varying class labels assigned by different participants.
The \textit{walking} class is the most consistently annotated class across participants and annotation methods, although it may not represent the maximum amount of labeled data points. Notably, the \textit{void} class, which is not visible in Figure \ref{fig:class_distribution_lin}, accounts for a substantial portion of the labeled data, ranging from 80\% to 96\%, depending on the annotation method employed.

Table \ref{tab:class_distribution} shows that daily activities that do not require extensive planning and are inherent to most people's everyday lives tend to be the most consistently annotated classes. Among these are activities such as \textit{walking}, \textit{cycling}, \textit{car\_driving}, or \textit{cooking}. On the other hand, activities like \textit{badminton}, \textit{weightlifting}, \textit{manual\_work}, and others are highly subject-dependent and occur only sporadically. The observation that the \textit{desk\_work} class exhibits the highest frequency is valid solely when employing the \circled{3} diary- or the \circled{4} GUI-methodology for data collection, suggesting a potential limitation or bias associated with this particular approach.
The classes pertaining to physical activities such as \textit{running, bus\_driving, yoga, badminton, weightlifting}, and \textit{sport} exhibit a distinct pattern of clustering within specific weeks, indicating a temporal dependency. This observation highlights the inherent bias introduced by the real-world recording environment in which the dataset was collected, potentially limiting the generalizability of the model to broader contexts.
Furthermore, it is crucial to note that the size of the \textit{void} class for the recall methods \circled{3} and \circled{4} is up to 16\% smaller than the in situ methods \circled{1} and \circled{2}.
While this disparity in class representation highlights the inherent complexities and challenges associated with the data collection and annotation procedures, it simultaneously presents an opportunity to address real-world imbalances and biases. By critically examining and accounting for these factors, the resulting models can potentially enhance their generalizability and applicability across diverse scenarios, ultimately contributing to a deeper understanding of the underlying phenomena under investigation.
\begin{table}\centering
\caption{This table presents a comprehensive overview of the number of data points for each activity class, categorized according to the different annotation methods employed during two distinct weeks. The columns are divided into two main sections, representing Week 1 and Week 2 of the data collection process. Within each week, the data points are further subdivided based on the annotation method used, labeled as button, app, diary, GUI (for Week 2 only).}\label{tab:class_distribution}
\scriptsize
\begin{adjustbox}{width=1\columnwidth}
    \begin{tabular}{lrrrrrrrrr}
\cellcolor[HTML]{b0b3b2} &\multicolumn{3}{c}{\cellcolor[HTML]{b0b3b2}\textbf{Week 1}} &\multicolumn{4}{c}{\cellcolor[HTML]{b0b3b2}\textbf{Week 2}} &\cellcolor[HTML]{b0b3b2}\textbf{Instances} \\
\cellcolor[HTML]{d4d4d4} &button &app &diary &button &app &diary &GUI &Total \\
\cellcolor[HTML]{d4d4d4}\textbf{laying} &0 &0 &135,000 &0 &0 &0 &0 &1 \\
\cellcolor[HTML]{d4d4d4}\textbf{sitting} &0 &0 &292,500 &0 &0 &607,500 &264,041 &13 \\
\cellcolor[HTML]{d4d4d4}\textbf{walking} &1,409,177 &1,380,112 &2,462,923 &1,446,000 &1,168,525 &2,160,712 &2,228,337 &245 \\
\cellcolor[HTML]{d4d4d4}\textbf{running} &16,500 &74,600 &84,000 &0 &0 &0 &0 &4 \\
\cellcolor[HTML]{d4d4d4}\textbf{cycling} &573,000 &574,100 &647,880 &616,500 &330,000 &610,500 &784,224 &92 \\
\cellcolor[HTML]{d4d4d4}\textbf{bus\_driving} &0 &67,500 &15,000 &0 &0 &0 &0 &2 \\
\cellcolor[HTML]{d4d4d4}\textbf{car\_driving} &708,000 &659,950 &1,683,176 &340,500 &155,450 &1,025,985 &954,882 &111 \\
\cellcolor[HTML]{d4d4d4}\textbf{vacuum\_cleaning} &49,500 &66,000 &67,500 &0 &42,000 &60,000 &47,955 &10 \\
\cellcolor[HTML]{d4d4d4}\textbf{laundry} &0 &27,000 &30,000 &33,000 &76,500 &112,500 &106,316 &19 \\
\cellcolor[HTML]{d4d4d4}\textbf{cooking} &229,500 &45,632 &269,132 &214,500 &96,000 &382,500 &419,848 &44 \\
\cellcolor[HTML]{d4d4d4}\textbf{eating} &0 &0 &5,753 &0 &0 &75,000 &92,850 &5 \\
\cellcolor[HTML]{d4d4d4}\textbf{shopping} &0 &57,000 &60,000 &0 &0 &45,000 &0 &3 \\
\cellcolor[HTML]{d4d4d4}\textbf{showering} &112,500 &0 &225,000 &34,500 &0 &251,777 &165,721 &20 \\
\cellcolor[HTML]{d4d4d4}\textbf{yoga} &0 &0 &0 &0 &0 &22,500 &30,680 &2 \\
\cellcolor[HTML]{d4d4d4}\textbf{sport} &39,000 &46,088 &45,000 &0 &0 &0 &0 &3 \\
\cellcolor[HTML]{d4d4d4}\textbf{playing\_games} &0 &0 &648,926 &0 &0 &45,000 &100,771 &6 \\
\cellcolor[HTML]{d4d4d4}\textbf{desk\_work} &36,000 &0 &3,646,985 &0 &0 &4,537,832 &1,137,214 &21 \\
\cellcolor[HTML]{d4d4d4}\textbf{guitar\_playing} &49,500 &0 &172,500 &111,000 &67,060 &217,500 &234,355 &24 \\
\cellcolor[HTML]{d4d4d4}\textbf{gardening} &48,000 &53,125 &43,718 &69,000 &69,650 &375,000 &421,794 &16 \\
\cellcolor[HTML]{d4d4d4}\textbf{table\_tennis} &0 &17,875 &0 &190,500 &108,725 &105,000 &171,177 &29 \\
\cellcolor[HTML]{d4d4d4}\textbf{badminton} &0 &0 &0 &0 &234,000 &210,000 &252,725 &6 \\
\cellcolor[HTML]{d4d4d4}\textbf{horse\_riding} &0 &66,000 &199,500 &0 &519,000 &495,000 &502,742 &18 \\
\cellcolor[HTML]{d4d4d4}\textbf{cleaning} &70,500 &100,500 &240,000 &48,000 &40,500 &502,500 &215,034 &25 \\
\cellcolor[HTML]{d4d4d4}\textbf{reading} &0 &0 &0 &0 &0 &45,000 &0 &1 \\
\cellcolor[HTML]{d4d4d4}\textbf{weightlifting} &0 &0 &0 &84,000 &0 &60,000 &0 &2 \\
\cellcolor[HTML]{d4d4d4}\textbf{manual\_work} &0 &0 &0 &0 &0 &69,425 &157,587 &3 \\
\cellcolor[HTML]{d4d4d4}\textbf{dish\_washing} &0 &0 &0 &10,500 &21,000 &52,500 &66,246 &9 \\
\cellcolor[HTML]{d4d4d4}\textbf{void} &61,581,367 &61,685,562 &53,948,051 &60,240,422 &60,510,012 &51,369,691 &55,083,923 &539 \\
\end{tabular}
\end{adjustbox}
\end{table}

\subsection{Missing Annotations and Consistency Across Methods}\label{sec:missing_annotation}
Missing Annotations and the consistency of labels set over the course of one week varied greatly depending on the study participant. However, tendencies with regard to specific methods are observable.

We computed missing annotations by merging all available annotations from the various methods used (button, app, diary, GUI) into an artificial global ground truth. We then compared each individual annotation layer against this consolidated ground truth to identify any missing annotations, leveraging the collective information as a reference point.

Method \circled{1}, pressing the situ button on the smartwatch's case, was not consistently used by every participant. Furthermore, this method carries the risk that either setting one of the two markers (start or end) is forgotten. An annotation where one marker is missing becomes therefore obsolete.
\begin{figure}[!t]
     \centering
     \includegraphics[width=.72\textwidth]{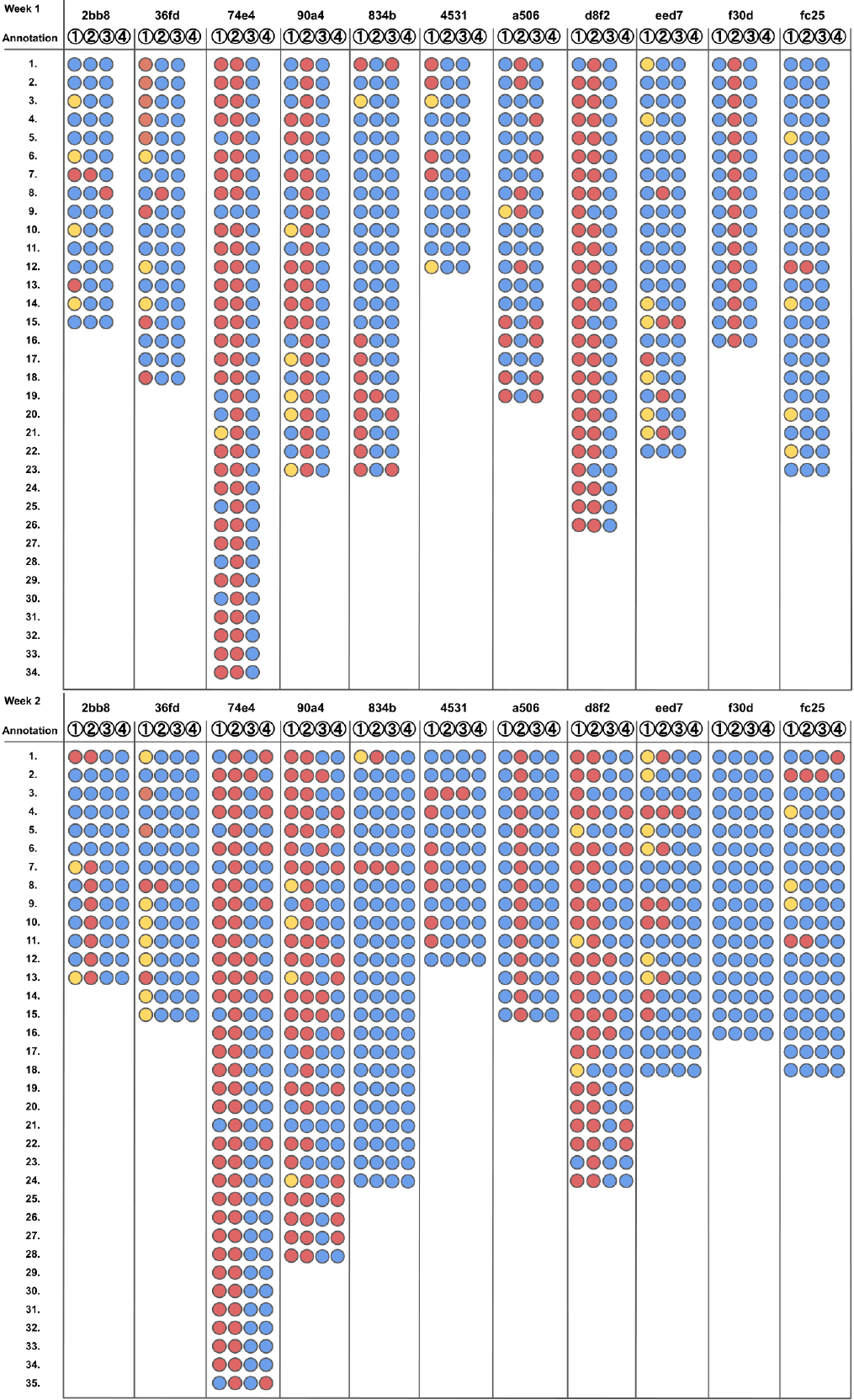}
     \caption[Comparing Annotation Methods: Missing Annotations and Consistency Across Methods.]{Missing annotations across all study participants and both weeks. The Y-axis shows the total number of annotations of one specific participant for the corresponding week. The color codes are as follows: \tikzcircle[fill=light-red]{3pt} Annotation is missing,  \tikzcircle[fill=light-yellow]{3pt} Annotation is partially missing (start or stop time), \tikzcircle[fill=light-blue]{3pt} Annotation is complete.

     The figure is inspired by \citep{brenner1999errors}, Figure 1.}
     \label{fig:consistency}
\end{figure}
The app-assisted annotation method \circled{2}, for which we used the app Strava, is well accepted among the participants who agreed with using third-party software. However, 4 participants, namely \textit{74e4}, \textit{90a4}, \textit{d8f2} and \textit{f30d} did not use the app continuously or refused to use it completely due to concerns regarding their private data.
Strava is a commercial app, that is freely available for download on the app stores, but it collects certain users' metadata. To label a time period with Strava, the participant needs to (1) take the smartphone, (2) open the app, (3) start a timer, set a label, and (4) end the timer. This procedure contains significantly more steps than other methods. Therefore, the average value of missing annotations results in 46.40\% (week 1) and 56.79\% (week 2). One participant found the annotation process in general very tedious and therefore dropped out of the study. These data have been excluded from the dataset and the evaluation.
Method \circled{3} \textit{pure self-recall}, writing an activity diary, got well accepted by every participant. As Figure \ref{fig:consistency} shows and the results in Table \ref{tab:avg_consistency} proof, it is overall the most complete annotation method with an average amount of missing annotations of 4.30 \% for the first and 8.14 \% for the second week.
\begin{table}[!htb]
 \caption[Comparing Annotation Methods: Missing annotations across all labeling methods (in \%) of both weeks.]{Missing annotations across all labeling methods (in \%) of both weeks. The columns contain the subject-ID of all participants. The last column shows the average percentage of missing annotations across every labeling method, for all participants.}
 \begin{adjustbox}{width=1\columnwidth}
 \begin{tabular}{lcccccccccccl}
 \hline
 \multicolumn{13}{|c|}{\multirow{2}{*}{\textbf{Week 1}}} \\
 \multicolumn{13}{|c|}{} \\ \hline
 \multicolumn{1}{|l|}{\textbf{Subject}} & \multicolumn{1}{l}{\textbf{2b88}} & \textbf{36fd} & \textbf{74e4} & \textbf{90a4} & \textbf{834b} & \textbf{4531} & \multicolumn{1}{l}{\textbf{a506}} & \multicolumn{1}{l}{\textbf{d8f2}} & \multicolumn{1}{l}{\textbf{eed7}} & \multicolumn{1}{l}{\textbf{f30d}} & \multicolumn{1}{l|}{\textbf{fc25}} & \multicolumn{1}{l|}{\textbf{Avg.}} \\ \hline
 \multicolumn{1}{|l|}{\textbf{\circled{1} \textit{in situ button}}} & 40 & 70.59 & 79.41 & 52.18 & 36.37 & 50 & 26.32 & 96.15 & 45.46 & 0 & \multicolumn{1}{c|}{26.81} & \multicolumn{1}{l|}{40.95} \\ \hline
 \multicolumn{1}{|l|}{\textbf{\circled{2} \textit{in situ app}}} & 13.30 & 5.89 & 97.06 & 100 & 5.00 & 0 & 36.84 & 92.30 & 22.73 & 100 & \multicolumn{1}{c|}{4.35} & \multicolumn{1}{l|}{43.40} \\ \hline
 \multicolumn{1}{|l|}{\textbf{\circled{3} \textit{pure self-recall}}} & 6.67 & 0 & 0 & 0 & 4.55 & 0 & 31.58 & 0 & 4.55 & 0 & \multicolumn{1}{c|}{0} & \multicolumn{1}{l|}{4.30} \\ \hline
 \multicolumn{13}{|c|}{\multirow{2}{*}{\textbf{Week 2}}} \\
 \multicolumn{13}{|c|}{} \\ \hline
 \multicolumn{1}{|l|}{\textbf{\circled{1} \textit{in situ button}}} & 23.08 & 73.33 & 92.00 & 82.14 & 8.33 & 76.47 & 0 & 95.33 & 61.11 & 0 & \multicolumn{1}{c|}{27.78} & \multicolumn{1}{l|}{49.05} \\ \hline
 \multicolumn{1}{|l|}{\textbf{\circled{2} \textit{in situ app}}} & 61.54 & 6.67 & 100 & 89.29 & 8.33 & 35.30 & 100 & 79.16 & 33.33 & 100 & \multicolumn{1}{c|}{11.11} & \multicolumn{1}{l|}{56.79} \\ \hline
 \multicolumn{1}{|l|}{\textbf{\circled{3} \textit{pure self-recall}}} & 0 & 0 & 8.57 & 17.88 & 4.17 & 35.30 & 0 & 12.50 & 5.56 & 0 & \multicolumn{1}{c|}{5.56} & \multicolumn{1}{l|}{8.14} \\ \hline
 \multicolumn{1}{|l|}{\textbf{\begin{tabular}[c]{@{}l@{}}\circled{4} \textit{time-series recall}\end{tabular}}} & 0 & 0 & 22.88 & 39.29 & 0 & 0 & 0 & 16.67 & 0 & 0 & \multicolumn{1}{c|}{5.56} & \multicolumn{1}{l|}{7.67} \\ \hline
 \end{tabular}
 \end{adjustbox}
 \label{tab:avg_consistency}
 \end{table}
By introducing the MAD-GUI, participants were able to inspect their daily data, get insights into what patterns of specific classes look like, and label them interactively. With an average amount of missing annotations of 7.67\%, this method became the most complete during the second study week.
 \begin{table}[!htb]
 \centering
 \caption[Comparing Annotation Methods: Average similarity between annotation methods according to the Cohan $\kappa$ score for both study weeks.]{Average similarity between annotation methods according to the Cohan $\kappa$ score for both study weeks. The columns are ordered subject-wise. The last column shows the average across all participants for one study week. The Direction column indicates in which the direction the Cohan $\kappa$ score is calculated and needs to be interpreted as follows: \circled{1} \textit{in situ button}, \circled{2} \textit{in situ app}, \circled{3} \textit{pure self-recall}, \circled{4} \textit{time-series recall}, (C/W) compared with.} 
 \begin{adjustbox}{width=.84\textwidth}
 \begin{tabular}{|l|r|r|r|r|r|r|r|r|r|r|r|r|l|}
 \hline
 \textbf{Week, Day} & \multicolumn{1}{c|}{\textbf{Direction}} & \multicolumn{1}{c|}{\textbf{2b88}} & \multicolumn{1}{c|}{\textbf{36fd}} & \multicolumn{1}{c|}{\textbf{4531}} & \multicolumn{1}{c|}{\textbf{74e4}} & \multicolumn{1}{c|}{\textbf{834b}} & \multicolumn{1}{c|}{\textbf{90a4}} & \multicolumn{1}{c|}{\textbf{a506}} & \multicolumn{1}{c|}{\textbf{d8f2}} & \multicolumn{1}{c|}{\textbf{eed7}} & \multicolumn{1}{c|}{\textbf{f30d}} & \multicolumn{1}{c|}{\textbf{fc25}} & \multicolumn{1}{c|}{\textbf{Avg.}} \\ \hline
 \textbf{1, 1} & \begin{tabular}[c]{@{}r@{}}\circled{3} C/W \circled{1}\\ \circled{3} C/W \circled{2}\end{tabular} & \begin{tabular}[c]{@{}r@{}}0.32\\ 0.69\end{tabular} & \begin{tabular}[c]{@{}r@{}}0.0\\ 0.85\end{tabular} & \begin{tabular}[c]{@{}r@{}}0.0\\ 0.69\end{tabular} & \begin{tabular}[c]{@{}r@{}}0.0\\ 0.0\end{tabular} & \begin{tabular}[c]{@{}r@{}}0.69\\ 0.76\end{tabular} & \begin{tabular}[c]{@{}r@{}}0.35\\ 0.0\end{tabular} & \begin{tabular}[c]{@{}r@{}}0.79\\ 0.0\end{tabular} & \begin{tabular}[c]{@{}r@{}}0.22\\ 0.0\end{tabular} & \begin{tabular}[c]{@{}r@{}}0.0\\ 0.90\end{tabular} & \begin{tabular}[c]{@{}r@{}}0.23\\ 0.0\end{tabular} & \begin{tabular}[c]{@{}r@{}}0.58\\ 0.49\end{tabular} &  \\ \hline
 \textbf{1, 2} & \begin{tabular}[c]{@{}r@{}}\circled{3} C/W \circled{1}\\ \circled{3} C/W \circled{2}\end{tabular} & \begin{tabular}[c]{@{}r@{}}0.64\\ 0.64\end{tabular} & \begin{tabular}[c]{@{}r@{}}0.69\\ 0.68\end{tabular} & \begin{tabular}[c]{@{}r@{}}0.0\\ 0.84\end{tabular} & \begin{tabular}[c]{@{}r@{}}0.09\\ 0.05\end{tabular} & \begin{tabular}[c]{@{}r@{}}0.85\\ 0.86\end{tabular} & \begin{tabular}[c]{@{}r@{}}0.05\\ 0.0\end{tabular} & \begin{tabular}[c]{@{}r@{}}0.51\\ 0.50\end{tabular} & \begin{tabular}[c]{@{}r@{}}0.0\\ 0.0\end{tabular} & \begin{tabular}[c]{@{}r@{}}0.55\\ 0.93\end{tabular} & \begin{tabular}[c]{@{}r@{}}0.47\\ 0.0\end{tabular} & \begin{tabular}[c]{@{}r@{}}0.74\\ 0.73\end{tabular} &  \\ \hline
 \textbf{1, 3} & \begin{tabular}[c]{@{}r@{}}\circled{3} C/W \circled{1}\\ \circled{3} C/W \circled{2}\end{tabular} & \begin{tabular}[c]{@{}r@{}}0.0\\ 0.86\end{tabular} & \begin{tabular}[c]{@{}r@{}}0.62\\ 0.0\end{tabular} & \begin{tabular}[c]{@{}r@{}}-0.03\\ -0.03\end{tabular} & \begin{tabular}[c]{@{}r@{}}0.0\\ 0.0\end{tabular} & \begin{tabular}[c]{@{}r@{}}0.39\\ 0.38\end{tabular} & \begin{tabular}[c]{@{}r@{}}0.56\\ 0.0\end{tabular} & \begin{tabular}[c]{@{}r@{}}0.53\\ 0.44\end{tabular} & \begin{tabular}[c]{@{}r@{}}0.0\\ 0.0\end{tabular} & \begin{tabular}[c]{@{}r@{}}0.05\\ 0.28\end{tabular} & \begin{tabular}[c]{@{}r@{}}0.53\\ 0.0\end{tabular} & \begin{tabular}[c]{@{}r@{}}0.51\\ 0.54\end{tabular} &  \\ \hline
 \textbf{1, 4} & \begin{tabular}[c]{@{}r@{}}\circled{3} C/W \circled{1}\\ \circled{3} C/W \circled{2}\end{tabular} & \begin{tabular}[c]{@{}r@{}}0.38\\ 0.99\end{tabular} & \begin{tabular}[c]{@{}r@{}}0.30\\ 0.69\end{tabular} & \begin{tabular}[c]{@{}r@{}}0.91\\ 0.90\end{tabular} & \begin{tabular}[c]{@{}r@{}}0.08\\ 0.0\end{tabular} & \begin{tabular}[c]{@{}r@{}}0.63\\ 0.74\end{tabular} & \begin{tabular}[c]{@{}r@{}}0.03\\ 0.0\end{tabular} & \begin{tabular}[c]{@{}r@{}}0.80\\ 0.57\end{tabular} & \begin{tabular}[c]{@{}r@{}}0.0\\ 0.69\end{tabular} & \begin{tabular}[c]{@{}r@{}}0.66\\ 0.80\end{tabular} & \begin{tabular}[c]{@{}r@{}}0.80\\ 0.0\end{tabular} & \begin{tabular}[c]{@{}r@{}}0.0\\ 0.0\end{tabular} &  \\ \hline
 \textbf{1, 5} & \begin{tabular}[c]{@{}r@{}}\circled{3} C/W \circled{1}\\ \circled{3} C/W \circled{2}\end{tabular} & \begin{tabular}[c]{@{}r@{}}0.33\\ 0.32\end{tabular} & \begin{tabular}[c]{@{}r@{}}0.33\\ 0.83\end{tabular} & \begin{tabular}[c]{@{}r@{}}0.0\\ 0.0\end{tabular} & \begin{tabular}[c]{@{}r@{}}0.04\\ 0.0\end{tabular} & \begin{tabular}[c]{@{}r@{}}0.39\\ 0.37\end{tabular} & \begin{tabular}[c]{@{}r@{}}0.32\\ 0.0\end{tabular} & \begin{tabular}[c]{@{}r@{}}0.93\\ 0.96\end{tabular} & \begin{tabular}[c]{@{}r@{}}0.0\\ 0.0\end{tabular} & \begin{tabular}[c]{@{}r@{}}-0.03\\ -0.31\end{tabular} & \begin{tabular}[c]{@{}r@{}}0.93\\ 0.0\end{tabular} & \begin{tabular}[c]{@{}r@{}}0.87\\ 0.89\end{tabular} &  \\ \hline
 \textbf{1, 6} & \begin{tabular}[c]{@{}r@{}}\circled{3} C/W \circled{1}\\ \circled{3} C/W \circled{2}\end{tabular} & \begin{tabular}[c]{@{}r@{}}0.0\\ -0.14\end{tabular} & \begin{tabular}[c]{@{}r@{}}0.0\\ 0.96\end{tabular} & \begin{tabular}[c]{@{}r@{}}0.75\\ 0.71\end{tabular} & \begin{tabular}[c]{@{}r@{}}0.07\\ 0.0\end{tabular} & \begin{tabular}[c]{@{}r@{}}0.0\\ 0.15\end{tabular} & \begin{tabular}[c]{@{}r@{}}0.34\\ 0.0\end{tabular} & \begin{tabular}[c]{@{}r@{}}0.67\\ -0.07\end{tabular} & \begin{tabular}[c]{@{}r@{}}0.0\\ 0.41\end{tabular} & \begin{tabular}[c]{@{}r@{}}0.42\\ 0.52\end{tabular} & \begin{tabular}[c]{@{}r@{}}0.99\\ 0.0\end{tabular} & \begin{tabular}[c]{@{}r@{}}0.84\\ 0.84\end{tabular} &  \\ \hline
 \textbf{1, 7} & \begin{tabular}[c]{@{}r@{}}\circled{3} C/W \circled{1}\\ \circled{3} C/W \circled{2}\end{tabular} & \begin{tabular}[c]{@{}r@{}}0.30\\ 0.78\end{tabular} & \begin{tabular}[c]{@{}r@{}}0.0\\ 0.15\end{tabular} & \begin{tabular}[c]{@{}r@{}}0.56\\ 0.69\end{tabular} & \begin{tabular}[c]{@{}r@{}}0.04\\ 0.0\end{tabular} & \begin{tabular}[c]{@{}r@{}}0.0\\ 0.10\end{tabular} & \begin{tabular}[c]{@{}r@{}}0.525\\ 0.0\end{tabular} & \begin{tabular}[c]{@{}r@{}}0.99\\ 0.43\end{tabular} & \begin{tabular}[c]{@{}r@{}}0.0\\ 0.0\end{tabular} & \begin{tabular}[c]{@{}r@{}}0.29\\ 0.42\end{tabular} & \begin{tabular}[c]{@{}r@{}}0.90\\ 0.0\end{tabular} & \begin{tabular}[c]{@{}r@{}}0.49\\ 0.77\end{tabular} &  \\ \hline
 \textbf{2, 1} & \begin{tabular}[c]{@{}r@{}}\circled{3} C/W \circled{1}\\ \circled{3} C/W \circled{2}\\ \circled{3} C/W \circled{4}\\ \circled{4} C/W  \circled{1}\\ \circled{4} C/W  \circled{2}\end{tabular} & \begin{tabular}[c]{@{}r@{}}0.30\\ 0.45\\ 0.85\\ 0.39\\ 0.53\end{tabular} & \begin{tabular}[c]{@{}r@{}}0.56\\ 0.77\\ 0.76\\ 0.43\\ 0.61\end{tabular} & \begin{tabular}[c]{@{}r@{}}0.36\\ 0.37\\ 0.56\\ 0.48\\ 0.45\end{tabular} & \begin{tabular}[c]{@{}r@{}}0.10\\ 0.0\\ 0.10\\ 0.43\\ 0.0\end{tabular} & \begin{tabular}[c]{@{}r@{}}0.51\\ 0.57\\ 0.48\\ 0.74\\ 0.70\end{tabular} & \begin{tabular}[c]{@{}r@{}}0.0\\ -0.02\\ 0.11\\ 0.0\\ 0.18\end{tabular} & \begin{tabular}[c]{@{}r@{}}0.88\\ 0.0\\ 0.90\\ 0.98\\ 0.0\end{tabular} & \begin{tabular}[c]{@{}r@{}}0.0\\ 0.63\\ 0.78\\ 0.0\\ 0.71\end{tabular} & \begin{tabular}[c]{@{}r@{}}0.41\\ 0.51\\ 0.74\\ 0.58\\ 0.57\end{tabular} & \begin{tabular}[c]{@{}r@{}}0.89\\ 0.0\\ 0.86\\ 0.82\\ 0.0\end{tabular} & \begin{tabular}[c]{@{}r@{}}0.85\\ 0.81\\ 0.46\\ 0.58\\ 0.55\end{tabular} &  \\ \hline
 \textbf{2, 2} & \begin{tabular}[c]{@{}r@{}}\circled{3} C/W \circled{1}\\ \circled{3}  C/W \circled{2}\\ \circled{3}  C/W \circled{4}\\ \circled{4} C/W  \circled{1}\\ \circled{4} C/W  \circled{2}\end{tabular} & \begin{tabular}[c]{@{}r@{}}0.82\\ 0.84\\ -0.02\\ -0.02\\ -0.02\end{tabular} & \begin{tabular}[c]{@{}r@{}}0.21\\ 0.75\\ 0.82\\ 0.11\\ 0.83\end{tabular} & \begin{tabular}[c]{@{}r@{}}0.91\\ 0.93\\ 0.62\\ 0.66\\ 0.63\end{tabular} & \begin{tabular}[c]{@{}r@{}}0.05\\ 0.0\\ 0.09\\ 0.38\\ 0.0\end{tabular} & \begin{tabular}[c]{@{}r@{}}0.47\\ 0.90\\ 0.84\\ 0.47\\ 0.92\end{tabular} & \begin{tabular}[c]{@{}r@{}}0.03\\ 0.0\\ 0.09\\ 0.77\\ 0.0\end{tabular} & \begin{tabular}[c]{@{}r@{}}0.70\\ 0.0\\ 0.70\\ 1.0\\ 0.0\end{tabular} & \begin{tabular}[c]{@{}r@{}}0.0\\ 0.87\\ 0.83\\ 0.0\\ 0.96\end{tabular} & \begin{tabular}[c]{@{}r@{}}0.29\\ 0.59\\ 0.56\\ 0.43\\ 0.71\end{tabular} & \begin{tabular}[c]{@{}r@{}}0.74\\ 0.0\\ 0.85\\ 0.70\\ 0.0\end{tabular} & \begin{tabular}[c]{@{}r@{}}0.36\\ 0.81\\ 0.46\\ 0.45\\ 0.46\end{tabular} &  \\ \hline
 \textbf{2, 3} & \begin{tabular}[c]{@{}r@{}}\circled{3} C/W \circled{1}\\ \circled{3}  C/W \circled{2}\\ \circled{3}  C/W \circled{4}\\ \circled{4} C/W  \circled{1}\\ \circled{4} C/W  \circled{2}\end{tabular} & \begin{tabular}[c]{@{}r@{}}0.70\\ 0.66\\ 0.68\\ 0.91\\ 0.74\end{tabular} & \begin{tabular}[c]{@{}r@{}}0.44\\ 0.44\\ 0.54\\ 0.53\\ 0.53\end{tabular} & \begin{tabular}[c]{@{}r@{}}0.0\\ 0.98\\ 0.62\\ 0.0\\ 0.82\end{tabular} & \begin{tabular}[c]{@{}r@{}}0.0\\ 0.0\\ 0.91\\ 0.0\\ 0.0\end{tabular} & \begin{tabular}[c]{@{}r@{}}0.62\\ 0.72\\ 0.49\\ 0.77\\ 0.74\end{tabular} & \begin{tabular}[c]{@{}r@{}}0.0\\ 0.0\\ -0.18\\ 0.0\\ 0.0\end{tabular} & \begin{tabular}[c]{@{}r@{}}0.90\\ 0.0\\ 0.90\\ 1.0\\ 0.0\end{tabular} & \begin{tabular}[c]{@{}r@{}}0.0\\ 0.47\\ 0.86\\ 0.0\\ 0.60\end{tabular} & \begin{tabular}[c]{@{}r@{}}0.28\\ 0.36\\ 0.39\\ 0.88\\ 0.79\end{tabular} & \begin{tabular}[c]{@{}r@{}}0.99\\ 0.0\\ 0.98\\ 0.96\\ 0.0\end{tabular} & \begin{tabular}[c]{@{}r@{}}0.68\\ 0.82\\ 0.58\\ 0.41\\ 0.65\end{tabular} &  \\ \hline
 \textbf{2, 4} & \begin{tabular}[c]{@{}r@{}}\circled{3} C/W \circled{1}\\ \circled{3}  C/W \circled{2}\\ \circled{3}  C/W \circled{4}\\ \circled{4} C/W  \circled{1}\\ \circled{4} C/W  \circled{2}\end{tabular} & \begin{tabular}[c]{@{}r@{}}0.45\\ 0.14\\ 0.17\\ 0.71\\ 0.82\end{tabular} & \begin{tabular}[c]{@{}r@{}}0.0\\ 0.85\\ 0.86\\ 0.0\\ 0.51\end{tabular} & \begin{tabular}[c]{@{}r@{}}0.83\\ 0.83\\ 0.84\\ 0.86\\ 0.86\end{tabular} & \begin{tabular}[c]{@{}r@{}}0.0\\ 0.0\\ 0.90\\ 0.0\\ 0.0\end{tabular} & \begin{tabular}[c]{@{}r@{}}-0.02\\ -0.02\\ -0.02\\ 0.80\\ 0.80\end{tabular} & \begin{tabular}[c]{@{}r@{}}0.05\\ 0.0\\ 0.0\\ 0.47\\ 0.0\end{tabular} & \begin{tabular}[c]{@{}r@{}}0.64\\ 0.0\\ 0.64\\ 1.0\\ 0.0\end{tabular} & \begin{tabular}[c]{@{}r@{}}0.0\\ 0.0\\ 0.84\\ 0.0\\ 0.0\end{tabular} & \begin{tabular}[c]{@{}r@{}}0.23\\ 0.26\\ 0.38\\ 0.68\\ 0.50\end{tabular} & \begin{tabular}[c]{@{}r@{}}0.67\\ 0.0\\ 0.86\\ 0.66\\ 0.0\end{tabular} & \begin{tabular}[c]{@{}r@{}}0.78\\ 0.87\\ 0.92\\ 0.76\\ 0.85\end{tabular} &  \\ \hline
 \textbf{2, 5} & \begin{tabular}[c]{@{}r@{}}\circled{3} C/W \circled{1}\\ \circled{3} C/W \circled{2}\\ \circled{3} C/W \circled{4}\\ \circled{4} C/W  \circled{1}\\ \circled{4} C/W  \circled{2}\end{tabular} & \begin{tabular}[c]{@{}r@{}}0.59\\ 0.0\\ 0.48\\ 0.5\\ 0.0\end{tabular} & \begin{tabular}[c]{@{}r@{}}0.0\\ 0.41\\ 0.47\\ 0.0\\ 0.34\end{tabular} & \begin{tabular}[c]{@{}r@{}}0.0\\ 0.77\\ 0.76\\ 0.0\\ 0.89\end{tabular} & \begin{tabular}[c]{@{}r@{}}0.0\\ 0.0\\ 0.16\\ 0.0\\ 0.0\end{tabular} & \begin{tabular}[c]{@{}r@{}}0.28\\ 0.28\\ 0.26\\ 0.82\\ 0.83\end{tabular} & \begin{tabular}[c]{@{}r@{}}0.09\\ 0.01\\ 0.09\\ 0.94\\ 0.60\end{tabular} & \begin{tabular}[c]{@{}r@{}}0.60\\ 0.0\\ 0.59\\ 0.99\\ 0.0\end{tabular} & \begin{tabular}[c]{@{}r@{}}0.0\\ 0.54\\ 0.82\\ 0.0\\ 0.49\end{tabular} & \begin{tabular}[c]{@{}r@{}}0.0\\ 0.54\\ 0.40\\ 0.0\\ 0.70\end{tabular} & \begin{tabular}[c]{@{}r@{}}0.73\\ 0.0\\ 0.43\\ 0.38\\ 0.0\end{tabular} & \begin{tabular}[c]{@{}r@{}}0.95\\ 0.92\\ 0.47\\ 0.46\\ 0.44\end{tabular} &  \\ \hline
 \textbf{2, 6} & \begin{tabular}[c]{@{}r@{}}\circled{3} C/W \circled{1}\\ \circled{3} C/W \circled{2}\\ \circled{3} C/W \circled{4}\\ \circled{4} C/W  \circled{1}\\ \circled{4} C/W  \circled{2}\end{tabular} & \begin{tabular}[c]{@{}r@{}}0.48\\ 0.0\\ 0.47\\ 0.98\\ 0.0\end{tabular} & \begin{tabular}[c]{@{}r@{}}0.0\\ 0.85\\ 0.86\\ 0.0\\ 0.88\end{tabular} & \begin{tabular}[c]{@{}r@{}}0.90\\ 0.96\\ 0.92\\ 0.95\\ 0.89\end{tabular} & \begin{tabular}[c]{@{}r@{}}0.0\\ 0.0\\ 0.77\\ 0.0\\ 0.0\end{tabular} & \begin{tabular}[c]{@{}r@{}}0.39\\ 0.39\\ 0.30\\ 0.47\\ 0.62\end{tabular} & \begin{tabular}[c]{@{}r@{}}0.0\\ 0.02\\ 0.0\\ 0.0\\ 0.69\end{tabular} & \begin{tabular}[c]{@{}r@{}}0.73\\ 0.0\\ 0.72\\ 0.99\\ 0.0\end{tabular} & \begin{tabular}[c]{@{}r@{}}0.0\\ 0.55\\ 0.83\\ 0.0\\ 0.43\end{tabular} & \begin{tabular}[c]{@{}r@{}}0.0\\ 0.83\\ 0.86\\ 0.0\\ 0.95\end{tabular} & \begin{tabular}[c]{@{}r@{}}0.20\\ 0.0\\ 0.74\\ 0.46\\ 0.0\end{tabular} & \begin{tabular}[c]{@{}r@{}}0.86\\ 0.87\\ 0.86\\ 0.83\\ 0.82\end{tabular} &  \\ \hline
 \textbf{2, 7} & \begin{tabular}[c]{@{}r@{}}\circled{3} C/W \circled{1}\\ \circled{3} C/W \circled{2}\\ \circled{3} C/W \circled{4}\\ \circled{4} C/W  \circled{1}\\ \circled{4} C/W  \circled{2}\end{tabular} & \begin{tabular}[c]{@{}r@{}}0.0\\ 0.0\\ 0.96\\ 0.0\\ 0.0\end{tabular} & \begin{tabular}[c]{@{}r@{}}0.0\\ 0.41\\ 0.47\\ 0.0\\ 0.80\end{tabular} & \begin{tabular}[c]{@{}r@{}}0.0\\ 0.76\\ 0.72\\ 0.0\\ 0.93\end{tabular} & \begin{tabular}[c]{@{}r@{}}0.0\\ 0.0\\ 0.0\\ 0.0\\ 0.0\end{tabular} & \begin{tabular}[c]{@{}r@{}}0.30\\ 0.30\\ 0.24\\ 0.79\\ 0.78\end{tabular} & \begin{tabular}[c]{@{}r@{}}0.0\\ 0.0\\ -0.01\\ 0.0\\ 0.0\end{tabular} & \begin{tabular}[c]{@{}r@{}}0.73\\ 0.0\\ 0.42\\ 0.67\\ 0.0\end{tabular} & \begin{tabular}[c]{@{}r@{}}0.40\\ 0.0\\ 0.79\\ 0.46\\ 0.0\end{tabular} & \begin{tabular}[c]{@{}r@{}}0.43\\ 0.14\\ 0.33\\ 0.71\\ 0.46\end{tabular} & \begin{tabular}[c]{@{}r@{}}0.91\\ 0.0\\ 0.92\\ 0.94\\ 0.0\end{tabular} & \begin{tabular}[c]{@{}r@{}}0.86\\ 0.86\\ 0.84\\ 0.78\\ 0.78\end{tabular} &  \\ \hline
 \textbf{Avg. Week 2} & \begin{tabular}[c]{@{}r@{}}\circled{3} C/W \circled{1}\\ \circled{3} C/W \circled{2}\\ \circled{3} C/W \circled{4}\\ \circled{4} C/W  \circled{1}\\ \circled{4} C/W  \circled{2}\end{tabular} & \textbf{\begin{tabular}[c]{@{}r@{}}0.48\\ 0.30\\ 0.51\\ 0.50\\ 0.30\end{tabular}} & \textbf{\begin{tabular}[c]{@{}r@{}}0.17\\ 0.64\\ 0.68\\ 0.15\\ 0.64\end{tabular}} & \textbf{\begin{tabular}[c]{@{}r@{}}0.43\\ 0.80\\ 0.72\\ 0.42\\ 0.78\end{tabular}} & \textbf{\begin{tabular}[c]{@{}r@{}}0.02\\ 0.0\\ 0.12\\ 0.42\\ 0.0\end{tabular}} & \textbf{\begin{tabular}[c]{@{}r@{}}0.36\\ 0.45\\ 0.37\\ 0.69\\ 0.78\end{tabular}} & \textbf{\begin{tabular}[c]{@{}r@{}}0.02\\ 0.0\\ 0.01\\ 0.31\\ 0.21\end{tabular}} & \textbf{\begin{tabular}[c]{@{}r@{}}0.74\\ 0.0\\ 0.70\\ 0.94\\ 0.0\end{tabular}} & \textbf{\begin{tabular}[c]{@{}r@{}}0.0\\ 0.44\\ 0.82\\ 0.07\\ 0.46\end{tabular}} & \textbf{\begin{tabular}[c]{@{}r@{}}0.23\\ 0.46\\ 0.33\\ 0.47\\ 0.67\end{tabular}} & \textbf{\begin{tabular}[c]{@{}r@{}}0.73\\ 0.0\\ 0.81\\ 0.70\\ 0.0\end{tabular}} & \textbf{\begin{tabular}[c]{@{}r@{}}0.76\\ 0.85\\ 0.66\\ 0.61\\ 0.65\end{tabular}} & \textbf{\begin{tabular}[c]{@{}l@{}}0.39\\ 0.36\\ 0.52\\ 0.48\\ 0.41\end{tabular}} \\ \hline
 \end{tabular}
 \end{adjustbox}
 \label{tab:cohen_kappa}
 \end{table}
Table \ref{tab:cohen_kappa} shows the resulting Cohen $\kappa$ scores. Due to the constraint that only one labeling method can be compared to a second one and since, according to Table \ref{tab:avg_consistency}, the most consistent annotation methods are \circled{3} \textit{pure self-recall} and \circled{4} \textit{time-series recall}, we used these methods as our baseline and compared them with every other method used in the study.
The second column indicates the comparison direction. The abbreviations used in this column are defined as follows: (\circled{3} C/W  \circled{1}) \textit{pure self-recall} compared with \textit{in situ button}, (\circled{3} C/W  \circled{2}) \textit{pure self-recall} compared with \textit{in situ app} and (\circled{3} C/W  \circled{4}) \textit{pure self-recall} compared with \textit{time-series recall}.
The direction (\circled{4} C/W  \circled{3}) is not explicitly included since Cohens $\kappa$ is bidirectional and both directions result in the same score. The score indicates how similar two annotators, or in our study labeling methods, are to each other. The resulting score is a decimal value between -1.0 and 1.0, where -1.0 means that the two annotators differ at most and 1.0 means complete similarity. 0.0 denotes that the target method was not used on that specific day.

Comparing the \circled{3} \textit{pure self-recall} method with the \circled{1} \textit{in situ button} and \circled{2} \textit{in situ app} method we can see that the final results for weeks 1 and 2 are proximate to one another. \circled{3} \textit{Pure self-recall} compared with the \circled{4} \textit{time-series recall} results in the highest similarity of 0.52. The comparison between the \circled{4} \textit{time-series recall} and the \circled{1} \textit{in situ button} as well as the \circled{2} \textit{in situ app} assisted annotations result in higher similarity than the prior comparison of \circled{3} \textit{pure self-recall} vs. both methods \circled{1} and \circled{2}. This means that subjects rather agree to the timestamps of the \textit{in situ} methods than to a self-written activity diary as soon as they can visually inspect the accelerometer data.
\subsection{Visual Time-Series Analysis}
\begin{figure*}[!b]
     \centering
     \includegraphics[width=\textwidth]{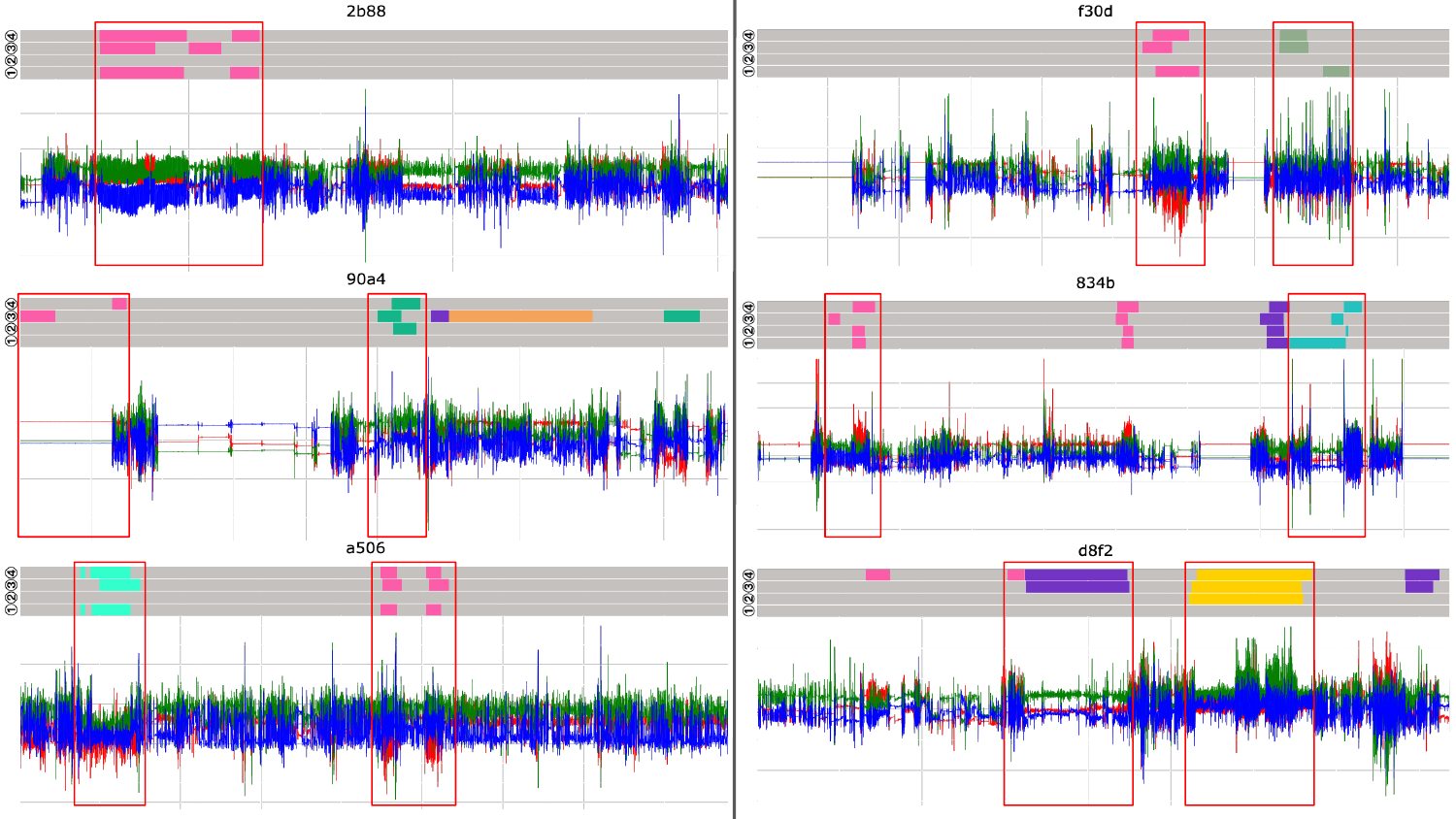}
     \caption[Comparing Annotation Methods: Visualization of participants' accelerometer data.]{Visualization of participants' accelerometer data on the sixth day in the second week of the study, together with annotations set by them. The four layers in the upper part of every participant's daily data represent the four annotation methods. The order is from bottom to top: \circled{1} \textit{in situ button}, \circled{2} \textit{in situ app}, \circled{3} \textit{pure self-recall} and \circled{4} \textit{time-series recall}.}
     \label{fig:label_comparison}
 \end{figure*}

Figure \ref{fig:label_comparison} shows exemplary the time-series of the sixth day of every participant's second week.
The four bars that are visible above the accelerometer data are the labels set by the participants for every layer.
The order is from bottom to top: \circled{1} \textit{in situ button}, \circled{2} \textit{in situ app}. \circled{3} \textit{pure self-recall}, and \circled{4} \textit{time-series recall}.
Examples of labels that differ with regard to the applied labeling method are marked with red boxes. The x-axis of every subplot represents roughly 8-9 hours of data. Most of the day was not labeled and is therefore categorized as \textit{void}. However, such long periods often contain shorter periods of other activities, like \textit{walking}. This makes it difficult to define a distinguishable \textit{void}-class, which results in false positive classifications of non-void samples. Figure \ref{fig:label_comparison} visually shows that each participant labels his or her data very subjectively. The long green-labeled periods of participant \textit{74e4} represent the class \textit{desk\_work}. The only other participant that used this label is \textit{90a4}.
Since each of the study participants works in an office environment and thus conclusively works at a desk, we can assume that the same class is classified as \textit{void} for all other study participants. This intra-class and inter-participant discrepancy becomes a problem whenever a model is trained that is supposed to generalize across individuals. To reduce these side effects and focus on the experiment itself, we decided to evaluate personalized models that take weekly data from participants into account.

The \textit{in situ button} annotation is empty for 5 participants: \textit{eed7}, \textit{36fd}, \textit{74e4}, \textit{90a4} and \textit{d8f2}.
Labels are only partially set or missing entirely for this annotation method and we therefore assume that participants tend to forget to press the button on the smartwatch. Both tables, \ref{tab:avg_consistency} and \ref{tab:cohen_kappa}, support this assumption, as this labeling method shows a high percentage of missing annotations as well as a low Cohen $\kappa$ score of 0.36\% (week 1) and 0.39\% (week 2). The \textit{pure self-recall} method \circled{3}, visible on the 2nd upper layer, is often misaligned compared to the in situ methods as well as the \textit{time-series recall} method \circled{4}. Participants tend to round up or down the start- and stop-time in steps of 5 or 10 minutes. For example, the annotations in Figure \ref{fig:label_comparison} given by the subjects \textit{2b88}, \textit{834b} or \textit{f30d}, show such incorrectly annotated data. The pink color represents the class \textit{walking}. With a closer look at the corresponding time-series data, one can see that the \textit{in situ button} annotation (bottom layer) and \textit{time-series recall} annotation (top layer) belongs to the typical periodic pattern of walking than the period labeled by \textit{pure self-recall}.

A consistent reliable performance in all labeling methods can only be observed at the participants \textit{4531} and \textit{fc25}. Other participants like \textit{eed7}, \textit{36fd}, \textit{74e4} or \textit{a506} are very precise in their annotations across methods, but are missing at least one layer of labels. The complete collection of visualizations is available in our dataset repository\footnote{\url{https://doi.org/10.5281/zenodo.7654684}}
\subsection{Effects on Classification}
\begin{figure}[!htb]
     \centering
     \includegraphics[width=0.7\columnwidth]{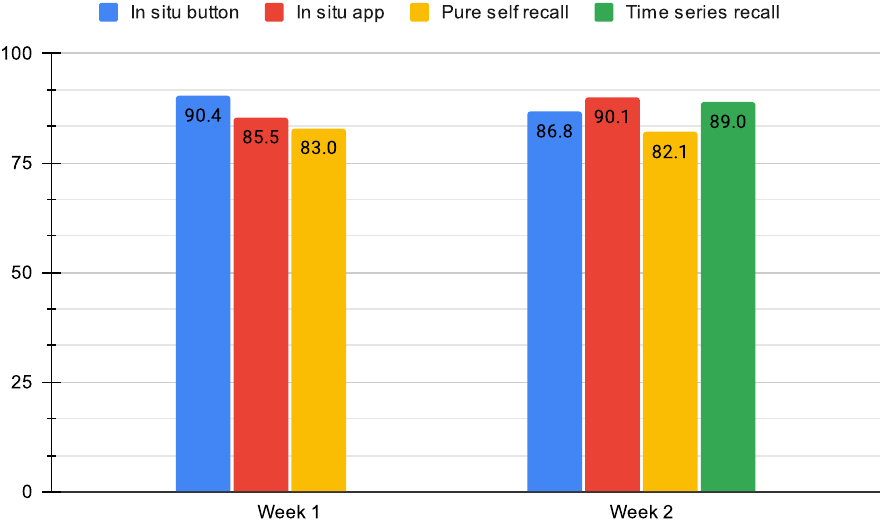}
     \caption[Comparing Annotation Methods: Deep Learning Evaluation.]{The overall mean F1-Scores for the Leave-One-Day-Out Cross Validation across all participants.
     In the first week, the participants used methods \circled{1} - \circled{3}. In the second week, we introduced method \circled{4}.}
     \label{fig:final_f1}
 \end{figure}

The results of our deep learning evaluation\footnote{Detailed results for every participant included in our deep learning evaluation can be accessed online on the Weights \& Biases platform: \url{https://tinyurl.com/4vxvfaed}.} suggest that the annotation method chosen can have a crucial impact on the classification ability of a trained deep learning model. Depending on the chosen methodology, the average F1-Score results differ by up to 8\%, as depicted in Figure \ref{fig:final_f1}.
In the first week, the in situ methodologies, button \circled{1} and app \circled{2}, generally perform better than the \textit{pure self-recall} diary \circled{3}.
Study participants mostly correctly estimated the duration of an activity, but tended to round up or down the start and end times. The in situ methods are up to 8\% better than the \textit{pure self-recall}, although the amount of annotated data available, due to missing annotations, is significantly lower than for other methods.
Although, we work with a dataset recorded in-the-wild, the deep learning results generally show a high F1-Score. This is untypical for such datasets but can be explained by the fact that the majority of the daily data are assigned to the \textit{void} class. This leaves proportionally only a few samples that are crucial for determining the classification performance.

Even though the number of available annotations that have been labeled by the study participants using the \textit{time-series recall} method \circled{4} is significantly higher with 92.33\%, the average F1-Score is 1.1\% lower (89.00\%) than the results reached with the App Assisted method (90.1\%). To understand this result it is crucial to look at Table \ref{tab:dl_results} in detail and take meta-information about the participants into account.
\begin{table}[!b]
 \centering
 \caption[Comparing Annotation Methods: In detail representation of the final F1-Scores for every annotation methodology and a week per study participant.]{In detail representation of the final F1-Scores for every annotation methodology and a week per study participant. The average F1-Scores are graphically visualized in Figure \ref{fig:final_f1}.}
 \label{tab:dl_results}
 \begin{adjustbox}{width=.8\textwidth}
 \begin{tabular}{|lccccccr|}
 \hline
 \multicolumn{8}{|c|}{\multirow{2}{*}{\textbf{Week 1}}} \\
 \multicolumn{8}{|c|}{} \\ \hline
 \multicolumn{1}{|l|}{\textbf{Subject}} & \multicolumn{1}{l|}{\textbf{2b88}} & \multicolumn{1}{l|}{\textbf{a506}} & \multicolumn{1}{l|}{\textbf{eed7}} & \multicolumn{1}{l|}{\textbf{fc25}} & \multicolumn{1}{l|}{\textbf{4531}} & \multicolumn{1}{l|}{\textbf{834b}} & \multicolumn{1}{l|}{\textbf{Average}} \\ \hline
 \multicolumn{1}{|l|}{\textbf{\circled{1} \textit{in situ button}}} & \multicolumn{1}{c|}{0,91} & \multicolumn{1}{c|}{0,92} & \multicolumn{1}{c|}{0,89} & \multicolumn{1}{c|}{0,89} & \multicolumn{1}{c|}{0,91} & \multicolumn{1}{c|}{0,91} & 90,4 \\ \hline
 \multicolumn{1}{|l|}{\textbf{\circled{2} \textit{in situ app}}} & \multicolumn{1}{c|}{0,92} & \multicolumn{1}{c|}{0,60} & \multicolumn{1}{c|}{0,91} & \multicolumn{1}{c|}{0,84} & \multicolumn{1}{c|}{0,93} & \multicolumn{1}{c|}{0,92} & 85,5 \\ \hline
 \multicolumn{1}{|l|}{\textbf{\circled{3} \textit{pure self-recall}}} & \multicolumn{1}{c|}{0,78} & \multicolumn{1}{c|}{0,76} & \multicolumn{1}{c|}{0,83} & \multicolumn{1}{c|}{0,86} & \multicolumn{1}{c|}{0,86} & \multicolumn{1}{c|}{0,89} & 83,0 \\ \hline
 \multicolumn{8}{|c|}{\multirow{2}{*}{\textbf{Week 2}}} \\
 \multicolumn{8}{|c|}{} \\ \hline
 \multicolumn{1}{|l|}{\textbf{\circled{1} \textit{in situ button}}} & \multicolumn{1}{c|}{0,88} & \multicolumn{1}{c|}{0,92} & \multicolumn{1}{c|}{0,90} & \multicolumn{1}{c|}{0,92} & \multicolumn{1}{c|}{0,86} & \multicolumn{1}{c|}{0,72} & 86,8 \\ \hline
 \multicolumn{1}{|l|}{\textbf{\circled{2} \textit{in situ app}}} & \multicolumn{1}{c|}{0,91} & \multicolumn{1}{c|}{na} & \multicolumn{1}{c|}{0,92} & \multicolumn{1}{c|}{0,94} & \multicolumn{1}{c|}{0,85} & \multicolumn{1}{c|}{0,88} & 90,1 \\ \hline
 \multicolumn{1}{|l|}{\textbf{\circled{3} \textit{pure self-recall}}} & \multicolumn{1}{c|}{0,81} & \multicolumn{1}{c|}{0,86} & \multicolumn{1}{c|}{0,82} & \multicolumn{1}{c|}{0,94} & \multicolumn{1}{c|}{0,83} & \multicolumn{1}{c|}{0,67} & 82,1 \\ \hline
 \multicolumn{1}{|l|}{\textbf{\circled{4} \textit{time-series recall}}} & \multicolumn{1}{c|}{0,90} & \multicolumn{1}{c|}{0,91} & \multicolumn{1}{c|}{0,95} & \multicolumn{1}{c|}{0,86} & \multicolumn{1}{c|}{0,86} & \multicolumn{1}{c|}{0,86} & 89,0 \\ \hline
 \end{tabular}
 \end{adjustbox}
 \end{table}
The participants mostly used their diary as a mnemonic aid for the graphical annotation method and tried to identify the corresponding periods in the acceleration data. The results of subjects \textit{2b88}, \textit{a506} and \textit{eed7} show that the performance of the classifier could be increased with graphical assistance. However, the F1-Score of \textit{2b88} is 0.01\% below the F1-Score of the \textit{in situ app} assisted annotation method \circled{2}.
These subjects have in common that they are already trained in interpreting acceleration data due to their prior knowledge and thus assign samples to specific classes more precisely.

\begin{figure}[!h]
     \centering
     \includegraphics[width=\columnwidth]{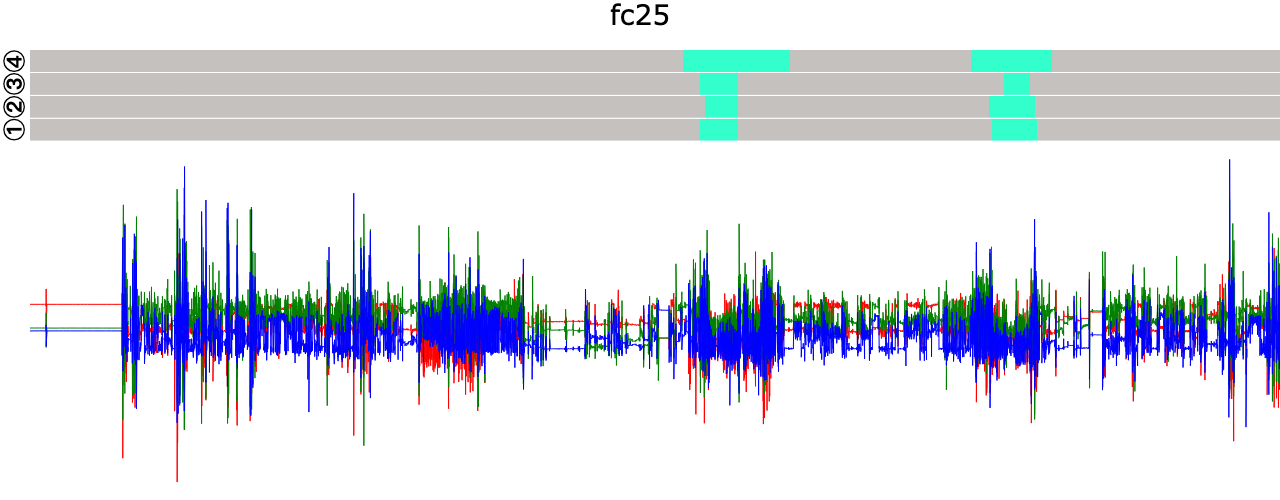}
     \caption[Comparing Annotation Methods: Visualization of the 1st day in week 2 of subject \textit{fc25}.]{Visualization of the 1st day in week 2 of subject \textit{fc25}. Differences can be seen in the upper annotation layer (\circled{4} \textit{time-series recall}), exhibiting larger differences regarding the annotated start- and stop times compared to other methods.}
     \label{fig:fc25}
 \end{figure}

Subjects \textit{fc25}, \textit{4531}, and \textit{834b}, on the other hand, do not have prior knowledge. Apart from subject \textit{834b}, the deep learning results show that presenting a visualization to an untrained participant rather harmed than helped the classifier.
If one looks at the visualizations of day 1 \& 6, week 2 of \textit{fc25}, see Figure \ref{fig:label_comparison} and \ref{fig:fc25}, the labels set by the subject with the help of the graphical interface, it is comprehensible that this study participant tended to be rather confused by the graphical representation and therefore labeled the data incorrectly.
\section{Discussion}
In our 2-week long-term study, we recorded the acceleration data of 11 participants using a smartwatch and analyzed it visually, statistically, and using deep learning. The findings of the visual and statistical analysis were confirmed by the deep learning result. They show that the underlying annotation procedure is crucial for the quality of the annotations and the success of the deep learning model.

The \textit{in situ button} method \circled{1} offers accuracy but brings the risk that the setting of a label is forgotten entirely or incompletely set. However, this method can be combined with additional on-device feedback or a smartphone app, so that greater accuracy and consistency of the annotation can be achieved. This involves a considerable implementation effort, which many scientists avoid because such projects, although of their significant value to the community, attract little attention in the scientific world. The use of existing, but often commercial, software and hardware is all too often accompanied by a loss of privacy. As our research has shown in passing, many users therefore shy away from using such products.

Through our investigation of the consistency of annotations between methodologies, we were able to show that participants in our study seem to prefer to write an activity diary (\textit{pure self-recall} method \circled{3}). This finding corresponds to what \citep{vaizman2018extrasensory} already points out. However, this method has the disadvantage that it can be imprecise, which is evident in the visualization of the data and annotations. Similarly, the activity diary methodology performed the least reliably among all methodologies, which has been confirmed by the deep learning model. Since the deep learning results using the \textit{in situ app} annotations \circled{3} are almost similar to the results given by the \textit{time-series recall} \circled{4}, even though the number of labeled samples is lower, it raises the question if a smaller set of high-quality annotations is more valuable for a classifier than a larger set of annotated data that comes with imprecise labels. This could mean that in future works we can reduce the amount of necessary training samples drastically if a certain annotation quality can be assured. However, this needs to be confirmed by further investigations.

Some participants reported that they found the support provided by the visual representation of the data helpful. The resulting Cohen $\kappa$ scores strengthen this impression since the F1-Scores are much higher when we compare the \textit{time-series recall} with both in situ methods vs. the \textit{pure self-recall}. This indicates that as soon as the participants received a visual inspection tool, they tended to annotate data at similar time periods as through the in situ methods since they can easily identify periods of activity that roughly correspond to the execution time they remember.
Our participants reported similar preferences, which led us to the conclusion that a digital diary that includes data visualization could combine the benefits of both annotation methods.

However, the study also showed that participants can find it difficult to interpret the acceleration data correctly and thus set inaccurate annotations. As our trained models show, this also has a strong influence on the classification result. If such a tool is to be made available to study participants, it must be ensured that they have the necessary knowledge and tools to be able to interpret these data.
Thus, to ensure the success of future long-term and real-world activity recognition projects, prior training of the study participants regarding data interpretation is of crucial importance if a data visualization is supposed to be used.

Apart from trying to solve annotation difficulties during the annotation phase itself, we can also partially counter wrong or noisy classified data by using machine learning techniques like Bootstrapping, see  \citep{miu2015bootstrapping} or using a loss function that specifically tries to counteract this problem, such as \citep{natarajan2013learning, ma2019bayesian}. By using Bootstrapping, the machine or deep learning classifier is initially trained by a small subset of high-confident labels and further improved by using additional data. However, this technique comes with the trade-off that whenever wrong-labeled data is introduced as training data, the error will get propagated into the model. This is an effect that sooner or later occurs as long as the annotation methodologies themselves are not further researched. Other machine learning techniques that can work with noisy labels, see \citep{song2022learning}, are already successfully tested for Computer Vision problems and can, in theory, be adopted for Human Activity Recognition. However, earlier research has shown that not every technique that is applicable in other fields is also applicable to sensor-based data. \citep{hoelzemann2020digging}.

\textbf{Cause of Missing Annotations:} We believe that specific activities are more likely to be forgotten during the labeling process than others. These activities are generally more spontaneous and require less dedicated preparation time. Examples might include classes like \textit{laying}, \textit{sitting}, \textit{walking}, \textit{bus\_driving}, \textit{car\_driving}, \textit{eating}, or \textit{desk\_work}. In contrast, other classes like \textit{shopping}, \textit{yoga}, \textit{playing\_games}, \textit{badminton}, \textit{cooking}, or \textit{horse\_riding} are often time-intensive, physically or mentally demanding, and frequently planned in advance or even take place at dedicated locations.  Therefore, it is likely that participants find these activities easier to recall and label accurately. Obtaining separate annotations for each activity through distinct and dedicated annotation processes would have yielded valuable insights; however, this approach was deemed unfeasible for the participants involved in our study. The immense time commitment and laborious efforts required from our participants to annotate each activity individually would have imposed an unreasonable burden, rendering such a comprehensive annotation strategy impractical within the constraints of our study.
\subsection{Discussing different Annotation Biases}
Directly quantifying the perceived workload of subjective tasks like data labeling is a complex challenge. This difficulty stems from several factors.
Firstly, individual differences in mental stamina and task perception mean what one person finds laborious, another might find manageable \citep{smith2019comparing}. Secondly, memory biases can lead to under- or overestimates of effort depending on the emotional context of the task or the participant's current state \citep{watkins2002implicit}. Social desirability bias can also come into play, with participants potentially downplaying their workload to appear competent or exaggerating it to justify breaks \citep{chung2003exploring}. Therefore, accurately quantifying the workload associated with each of the four labeling methods presents a significant challenge. While surveys, like the NASA TLX \citep{aeronauticsNASATaskLoad1986}, asking participants about perceived effort hold some value, these results are inherently subjective and can be heavily influenced by individual experiences and biases. While aiming for a fully objective measure of workload is desirable, it might require collecting more personal data from participants. This additional data could include details like preferred wearable devices (e.g., smartwatches), smartphone usage patterns, or individual memory recall capabilities or the emotional state of a participant \citep{ghosh2015annotation}. While recording and quantifying this type of personal data would have provided valuable insights, it  would have also significantly increased the workload placed on participants.  This additional workload fell outside the scope of the current study, which prioritized collecting data through the four predefined methods.

However, we need to acknowledge that several biases could have been introduced due to the chosen annotation guidelines and tools. For example, the usage of in situ annotation methods during the day can have a positive effect on the self-recall capabilities of a participant at the end of the day. The comparison of consistencies across methods does not confirm that this effect indeed occurred. Every study participant showed an almost complete overall profile of self-recall annotations, even though the person has not used or has incomplete in situ annotations, see Figure \ref{fig:consistency}. However, deeper investigations are needed to be able to understand such effects better.

\citep{yordanova2018ardous} lists the following 3 biases for sensor-based human activity data: Self-Recall bias \citep{valuri2005validity}, Behavior bias \citep{friesen2020all} and the Self-Annotation bias \citep{yordanova2018ardous}.
We showed that indeed a time-deviation bias (which can be seen as a self-recall bias) has been introduced to annotations created with the \textit{pure self-recall} method \circled{3}, and that such a bias affects the classifier negatively. However, visualizing the sensor data can counter this effect because it was easier for participants to detect active phases in hindsight.

A behavior bias can be neglected, because the participants were not monitored by a person or video camera during the day and the minimalistic setup of one wrist-worn smartwatch does not influence one's behavior since the wearing comfort of such a device is generally perceived as positive \citep{pal2020future}.
A self-annotation bias, a bias that occurs if the annotator labels their data in an isolated environment and cannot refer to an expert to verify an annotation, did occur as well. With the deep learning analysis, we were able to show that the classifier was less negatively impacted by this bias than by time-deviation bias.

A \textbf{Parallel annotation bias} can arise in two scenarios: when multiple annotators independently label the same data, or when a single annotator uses multiple labeling methods for the same data, where the application of one method influences the subsequent labeling decisions made with other methods.
There are three main ways this bias can manifest:
\begin{enumerate}
    \item Anchoring bias \citep{lieder2018anchoring}: The initial labeling method might act as an anchor, subtly influencing the annotator's decisions when using subsequent methods, even if their initial assessment might differ.
    \item Confirmation bias \citep{klayman1995varieties}: The annotator might subconsciously favor interpretations that align with labels generated from previous methods, overlooking alternative possibilities.
    \item Method bias \citep{min2016common}: Certain methods might inherently be easier or more difficult to use for specific types of data, potentially leading to systematic inconsistencies across the labeled data.
\end{enumerate}

The presence of parallel annotation bias in this context suggests that the annotations might not be entirely independent between methods, potentially impacting the overall quality of the data. Anchoring and confirmation bias can lead to a lack of diversity in annotations and potentially perpetuate errors. Method bias can introduce inconsistencies that complicate data analysis. We recognize the possibility of parallel annotation bias in our dataset, where applying one labeling method might influence subsequent methods used by the same participant. However, prioritizing participant engagement, we opted for a parallel approach. This decision ensured the workload remained manageable and prevented participant dropout from the study.

\section{Conclusions}
We argue that the annotation methodologies for benchmark datasets in Human Activity Recognition do not yet capture the attention it should. Data annotation is a laborious and time-consuming task that often cannot be performed accurately and conscientiously without the right tools. However, there is a very limited number of tools that can be used for this purpose and often they do not pass the prototype status.

Only a few scientific publications, such as \citep{reining2020annotation}, focus on annotations and their quality. However, the use of properly annotated data drastically affects the final capacities of the trained machine or deep learning model. Therefore, we consider our study to be important for the HAR community, as it analyzes this topic in greater depth and thus provides important insights that go beyond the current state of science. Table \ref{tab:conclusion} summarizes the advantages and disadvantages of every method.
 \begin{table}[!htb]
 \centering
 \caption[Comparing Annotation Methods: Comparison of advantages and disadvantages of all annotation methods used in this study.]{Comparison of advantages and disadvantages of all annotation methods used in this study.}
 \label{tab:conclusion}
 \resizebox{\textwidth}{!}{%
 \begin{tabular}{|l|l|l|l|}
 \hline
 \multicolumn{1}{|c|}{\textbf{Methodology}} & \multicolumn{1}{c|}{\textbf{Advantages}} & \multicolumn{1}{c|}{\textbf{Disadvantages}} \\ \hline
 \begin{tabular}[c]{@{}l@{}}\circled{1} \textit{in situ button}\end{tabular} & \begin{tabular}[c]{@{}l@{}}- Easy to implement and use\\- Can be improved with feedback mechanisms\end{tabular} & \begin{tabular}[c]{@{}l@{}}- Participants tend to forget pressing a button\\- Many incomplete annotations that become \\ \;\;unusable for the classifier\end{tabular} \\ \hline

 \begin{tabular}[c]{@{}l@{}}\circled{2} \textit{in situ app}\end{tabular} & \begin{tabular}[c]{@{}l@{}}- Tracking apps are already widely used and accepted,\\ \;\;therefore low acceptance threshold
 \\- Can be improved with feedback mechanisms or\\ \;\;additional smartphone functionalities\\- List of possible annotations can be expanded with \\ \;\;minimum effort\\- Participants tend to set very precise annotations.\end{tabular} & \begin{tabular}[c]{@{}l@{}}- Data and privacy concerns if a commercial app is used.\\- Participants often  forgot to set an annotation,\\ \;\;especially when they were unfamiliar with tracking apps.\\- Implementation workload may be very high\end{tabular} \\ \hline

 \begin{tabular}[c]{@{}l@{}}\circled{3} \textit{pure self-recall}\\\end{tabular} & \begin{tabular}[c]{@{}l@{}}- Easy to use even without technical knowledge\\ \;\;(a handwritten diary)\\- Most accepted method in our experiment\\- Annotations are very consistent\end{tabular} & \begin{tabular}[c]{@{}l@{}}- Can be very imprecise\\- Only suitable for coarse activity labels and activities\\ \;\;that were performed for long periods of time,\\ \;\;like \textit{walking} or \textit{running}\end{tabular} \\ \hline
 \begin{tabular}[c]{@{}l@{}}\circled{4} \textit{time-series} \\ recall\end{tabular} & \begin{tabular}[c]{@{}l@{}}- Visualization of data helps participants to set\\ \;\;annotations more accurate than using the pure\\ \;\;self-recall method \circled{3}\end{tabular}&
 \begin{tabular}[c]{@{}l@{}}- Available tools are often in the state of a prototype\\ \;\;and need additional developments and adjustments\\ \;\;and are therefore not impromptu usable.\\- Participants need to be trained to be able to interpret\\ \;\;sensor data.\end{tabular}  \\ \hline
 \end{tabular}%
 }
 \end{table}

High-quality annotations are crucial for accurate activity recognition, especially in uncontrolled real-world settings where video recordings are unavailable for ground truth verification. To address this challenge, further research on activity data annotation methodologies is necessary. These methodologies should empower annotators to label data in a way that comprehensively captures the subtleties of everyday life. The annotations must not only be extensive but also complete and coherent, ensuring a consistent and well-defined understanding for the AI model to learn from. Furthermore, leveraging learning methodologies like Weakly Supervised Learning methods, exemplified by works such as \citep{wang2019attention} and \citep{wang2021sequential}, can potentially utilize datasets like ours. However, a more comprehensive evaluation is needed to determine their suitability for real-world application.

The combination of a (handwritten) diary with a correction aided by a data visualization in hindsight shows the best results in terms of consistency and missing annotations and provides accurate start and end times. However, this combination results in additional work for the study participants and therefore, remains a trade-off between additional workload and annotation quality.

\textbf{Lesson Learned}.
During this study, we gained insights about the effects of different annotation methods on the reliability and consistency of annotations and finally on the classifier itself, but also about training deep learning models on data recorded in-the-wild. In this chapter, we would like to share these insights to help other researchers perform their experiments more successfully.
With regards to Table \ref{tab:conclusion}, we are able to narrow down specific study setups that either benefit more from self-recall or in situ annotation methods.
\begin{enumerate}
    \item Due to the good acceptance and the low workload for study participants we can recommend a self-recall method for studies where label precision is not the highest priority and rough estimations of activities are sufficient.
     \item According to our study, we can increase the label precision of the self-recall method with additional software that visualizes the raw data, e.g. \citep{ollenschlager2022mad}. We recommend considering the implementation of such a module and providing this software to participants together with an introduction on how to interpret sensor data. According to \citep{vaizman2018extrasensory} the self-recall method can also be effectively improved by introducing server guesses of activities or visually organizing the day chronologically.
     \item In situ annotations result generally in more precise labels. However, the label process is more labor intensive than a self-recall method, since it can take a lot of time and often includes many steps to set the label. We argue, that smaller studies with participants who agree with performing such laborious work can benefit from this method. Such a system needs to be implemented carefully and with a holistic concept in order to not be seen as a burden by the participants \citep{yordanova2019challenges}.
     \item Annotating data with commercial apps, like Strava, is negligible due to data and privacy concerns.
     \item In situ annotation can have the same benefits as an app solution. However, only if researchers have access to the programming interface of the recording device and can implement additional features that help participants not forget to set a label.
 \end{enumerate}
As part of our annotation guidelines, we allowed our study participants to name their activities as they wished. Therefore, we were forced to simplify certain activities.
To be able to create a real-world dataset that contains complex classes or even classes that consist of several subclasses, more elaborated annotation methods and tools must be developed. We believe that with the currently available resources, the hurdle lies very high for such datasets to be annotated accurately.

Our study includes people who cycle to work in their daily work routine and others who commute by public transport or work in a home office environment. Thus, each study participant has his or her set of daily repetitive activities.
Due to the nature of our dataset as one recorded in a real-world and long-term scenario, the number of labeled samples is rather small, and given labels vary participant-dependent. This mix of factors creates a bias in the dataset and we concluded that a cross-participant train-/test-strategy is not appropriate for our study design and would not give meaningful insights, since every study participant has their own set of unique activities which are too different and hardly generalizable. Therefore, for certain studies, the commonly known and accepted Leave-One-Subject-Out Cross-Validation is not suitable.
\section*{Conflict of Interest Statement}

The authors declare that the research was conducted in the absence of any commercial or financial relationships that could be construed as a potential conflict of interest.

\section*{Author Contributions}
\textbf{Alexander Hoelzemann} has performed the implementation and implemented all studies and visualizations, \textbf{Kristof Van Laerhoven} has guided this work and assisted in the methodologies. Both authors have contributed substantially to the writing of this manuscript.
\section*{Funding}
This project is funded by the Deutsche Forschungsgemeinschaft (DFG, German Research Foundation) – 425868829 and is part of Priority Program SPP2199 Scalable Interaction Paradigms for Pervasive Computing Environments.
\section*{Ethics}
The studies involving human participants were reviewed and approved by the Ethics Committee of the University of Siegen (ethics vote \#ER\textunderscore12\textunderscore2019). The participants provided their written informed consent to participate in this study and were briefed before the data capture about the study goals.
\section*{Consent to participate}
Informed consent was obtained from all individual participants included in the study.
\section*{Consent for publication}
The participants have consented to the submission of the manuscript to the journal.
\section*{Supplemental Data}
All code is made publicly available on our GitHub repository \url{https://github.com/ahoelzemann/annotationMatters}. Furthermore, a summary of all deep learning experiments can be accessed on the Weights and Biases \url{https://tinyurl.com/4vxvfaed}.
\section*{Data Availability Statement}
The datasets generated for this study can be found in the Zenodo Repository \url{https://doi.org/10.5281/zenodo.7654684}.
\bibliographystyle{unsrt}
\bibliography{references}

\begin{thebibliography}{10}

\bibitem{roggen2010collecting}
Daniel Roggen, Alberto Calatroni, Mirco Rossi, Thomas Holleczek, Kilian
  F{\"o}rster, Gerhard Tr{\"o}ster, Paul Lukowicz, David Bannach, Gerald Pirkl,
  Alois Ferscha, et~al.
\newblock Collecting complex activity datasets in highly rich networked sensor
  environments.
\newblock In {\em 2010 Seventh international conference on networked sensing
  systems (INSS)}, pages 233--240. IEEE, 2010.

\bibitem{mekruksavanich2021recognition}
Sakorn Mekruksavanich and Anuchit Jitpattanakul.
\newblock Recognition of real-life activities with smartphone sensors using
  deep learning approaches.
\newblock In {\em 2021 IEEE 12th International Conference on Software
  Engineering and Service Science (ICSESS)}, pages 243--246. IEEE, 2021.

\bibitem{friesen2020all}
Kenzie~B Friesen, Zhaotong Zhang, Patrick~G Monaghan, Gretchen~D Oliver, and
  Jaimie~A Roper.
\newblock All eyes on you: how researcher presence changes the way you walk.
\newblock {\em Scientific Reports}, 10(1):1--8, 2020.

\bibitem{zhao2013healthy}
Kunlun Zhao, Junzhao Du, Congqi Li, Chunlong Zhang, Hui Liu, and Chi Xu.
\newblock Healthy: A diary system based on activity recognition using
  smartphone.
\newblock In {\em 2013 IEEE 10th international conference on mobile Ad-Hoc and
  sensor systems}, pages 290--294. IEEE, 2013.

\bibitem{natarajan2013learning}
Nagarajan Natarajan, Inderjit~S Dhillon, Pradeep~K Ravikumar, and Ambuj Tewari.
\newblock Learning with noisy labels.
\newblock {\em Advances in neural information processing systems}, 26, 2013.

\bibitem{bock2022investigating}
Marius Bock, Alexander Hoelzemann, Michael Moeller, and Kristof Van~Laerhoven.
\newblock Investigating (re) current state-of-the-art in human activity
  recognition datasets.
\newblock {\em Frontiers in Computer Science}, 4, 2022.

\bibitem{akbari2021facilitating}
Ali Akbari, Jonathan Martinez, and Roozbeh Jafari.
\newblock Facilitating human activity data annotation via context-aware change
  detection on smartwatches.
\newblock {\em ACM Transactions on Embedded Computing Systems (TECS)},
  20(2):1--20, 2021.

\bibitem{cleland2014evaluation}
Ian Cleland, Manhyung Han, Chris Nugent, Hosung Lee, Sally McClean, Shuai
  Zhang, and Sungyoung Lee.
\newblock Evaluation of prompted annotation of activity data recorded from a
  smart phone.
\newblock {\em Sensors}, 14(9):15861--15879, 2014.

\bibitem{ollenschlager2022mad}
Malte Ollenschl{\"a}ger, Arne K{\"u}derle, Wolfgang Mehringer, Ann-Kristin
  Seifer, J{\"u}rgen Winkler, Heiko Ga{\ss}ner, Felix Kluge, and Bjoern~M
  Eskofier.
\newblock Mad gui: An open-source python package for annotation and analysis of
  time-series data.
\newblock {\em Sensors}, 22(15):5849, 2022.

\bibitem{bock2021improving}
Marius Bock, Alexander H{\"o}lzemann, Michael Moeller, and Kristof
  Van~Laerhoven.
\newblock Improving deep learning for har with shallow lstms.
\newblock In {\em 2021 International Symposium on Wearable Computers}, pages
  7--12, 2021.

\bibitem{ordonez_2016}
Francisco Ordóñez and Daniel Roggen.
\newblock Deep convolutional and lstm recurrent neural networks for multimodal
  wearable activity recognition.
\newblock {\em Sensors}, 16:115, 01 2016.

\bibitem{berlin2012detecting}
Eugen Berlin and Kristof Van~Laerhoven.
\newblock Detecting leisure activities with dense motif discovery.
\newblock In {\em Proceedings of the 2012 ACM Conference on Ubiquitous
  Computing}, pages 250--259, 2012.

\bibitem{vaizman2018extrasensory}
Yonatan Vaizman, Katherine Ellis, Gert Lanckriet, and Nadir Weibel.
\newblock Extrasensory app: Data collection in-the-wild with rich user
  interface to self-report behavior.
\newblock In {\em Proceedings of the 2018 CHI conference on human factors in
  computing systems}, pages 1--12, 2018.

\bibitem{thomaz2015practical}
Edison Thomaz, Irfan Essa, and Gregory~D Abowd.
\newblock A practical approach for recognizing eating moments with
  wrist-mounted inertial sensing.
\newblock In {\em Proceedings of the 2015 ACM international joint conference on
  pervasive and ubiquitous computing}, pages 1029--1040, 2015.

\bibitem{sztylerOnBodyLocalizationWearable2016}
Timo Sztyler and Heiner Stuckenschmidt.
\newblock On-{Body} {Localization} of {Wearable} {Devices}: {An}
  {Investigation} of {Position}-{Aware} {Activity} {Recognition}.
\newblock In {\em {IEEE} {International} {Conference} on {Pervasive}
  {Computing} and {Communications}}, pages 1--9, 2016.

\bibitem{gjoreski2018university}
Hristijan Gjoreski, Mathias Ciliberto, Lin Wang, Francisco Javier~Ordonez
  Morales, Sami Mekki, Stefan Valentin, and Daniel Roggen.
\newblock The university of sussex-huawei locomotion and transportation dataset
  for multimodal analytics with mobile devices.
\newblock {\em IEEE Access}, 6:42592--42604, 2018.

\bibitem{yordanova2018ardous}
Kristina~Y. Yordanova, Adeline Paiement, Max Schr{\"{o}}der, Emma Tonkin,
  Przemyslaw Woznowski, Carl~Magnus Olsson, Joseph Rafferty, and Timo Sztyler.
\newblock Challenges in annotation of user data for ubiquitous systems: Results
  from the 1st {ARDUOUS} workshop.
\newblock {\em CoRR}, abs/1803.05843, 2018.

\bibitem{hoelzemann2019using}
Alexander Hoelzemann, Henry Odoemelem, and Kristof Van~Laerhoven.
\newblock Using an in-ear wearable to annotate activity data across multiple
  inertial sensors.
\newblock In {\em Proceedings of the 1st International Workshop on Earable
  Computing}, pages 14--19, 2019.

\bibitem{stikic2011weakly}
Maja Stikic, Diane Larlus, Sandra Ebert, and Bernt Schiele.
\newblock Weakly supervised recognition of daily life activities with wearable
  sensors.
\newblock {\em IEEE transactions on pattern analysis and machine intelligence},
  33(12):2521--2537, 2011.

\bibitem{cruz2019semi}
Dagoberto Cruz-Sandoval, Jessica Beltran-Marquez, Matias Garcia-Constantino,
  Luis~A Gonzalez-Jasso, Jesus Favela, Irvin~Hussein Lopez-Nava, Ian Cleland,
  Andrew Ennis, Netzahualcoyotl Hernandez-Cruz, Joseph Rafferty, et~al.
\newblock Semi-automated data labeling for activity recognition in pervasive
  healthcare.
\newblock {\em Sensors}, 19(14):3035, 2019.

\bibitem{reining2020annotation}
Christopher Reining, Fernando~Moya Rueda, Friedrich Niemann, Gernot~A Fink, and
  Michael ten Hompel.
\newblock Annotation performance for multi-channel time series har dataset in
  logistics.
\newblock In {\em 2020 IEEE International Conference on Pervasive Computing and
  Communications Workshops (PerCom Workshops)}, pages 1--6. IEEE, 2020.

\bibitem{valuri2005validity}
Giulietta Valuri, Mark Stevenson, Caroline Finch, Peter Hamer, and Bruce
  Elliott.
\newblock The validity of a four week self-recall of sports injuries.
\newblock {\em Injury Prevention}, 11(3):135--137, 2005.

\bibitem{van2008using}
Kristof Van~Laerhoven, David Kilian, and Bernt Schiele.
\newblock Using rhythm awareness in long-term activity recognition.
\newblock In {\em 2008 12th IEEE International Symposium on Wearable
  Computers}, pages 63--66. IEEE, 2008.

\bibitem{tapia2004activity}
Emmanuel~Munguia Tapia, Stephen~S Intille, and Kent Larson.
\newblock Activity recognition in the home using simple and ubiquitous sensors.
\newblock In {\em International conference on pervasive computing}, pages
  158--175. Springer, 2004.

\bibitem{gjoreski2017versatile}
Hristijan Gjoreski, Mathias Ciliberto, Francisco Javier~Ordo{\~n}ez Morales,
  Daniel Roggen, Sami Mekki, and Stefan Valentin.
\newblock A versatile annotated dataset for multimodal locomotion analytics
  with mobile devices.
\newblock In {\em Proceedings of the 15th ACM Conference on Embedded Network
  Sensor Systems}, pages 1--2, 2017.

\bibitem{tonkin2018talk}
Emma~L Tonkin, Alison Burrows, Przemys{\l}aw~R Woznowski, Pawel Laskowski,
  Kristina~Y Yordanova, Niall Twomey, and Ian~J Craddock.
\newblock Talk, text, tag? understanding self-annotation of smart home data
  from a user’s perspective.
\newblock {\em Sensors}, 18(7):2365, 2018.

\bibitem{schroder2016tool}
Max Schr{\"o}der, Kristina Yordanova, Sebastian Bader, and Thomas Kirste.
\newblock Tool support for the online annotation of sensor data.
\newblock In {\em Proceedings of the 3rd International Workshop on Sensor-based
  Activity Recognition and Interaction}, pages 1--7, 2016.

\bibitem{leonardis2002multiple}
Ale{\v{s}} Leonardis, Horst Bischof, and Jasna Maver.
\newblock Multiple eigenspaces.
\newblock {\em Pattern recognition}, 35(11):2613--2627, 2002.

\bibitem{huynh2008human}
Duy T{\^a}m~Gilles Huynh.
\newblock Human activity recognition with wearable sensors.
\newblock {\em Technische Universit{\"a}t Darmstadt}, pages 59--65, 2008.

\bibitem{hassan2021autoact}
Iqbal Hassan, Abtahi Mursalin, Robin~Bin Salam, Nazmus Sakib, and HM~Zabir
  Haque.
\newblock Autoact: An auto labeling approach based on activities of daily
  living in the wild domain.
\newblock In {\em 2021 Joint 10th International Conference on Informatics,
  Electronics \& Vision (ICIEV) and 2021 5th International Conference on
  Imaging, Vision \& Pattern Recognition (icIVPR)}, pages 1--8. IEEE, 2021.

\bibitem{wu2022survey}
Xingjiao Wu, Luwei Xiao, Yixuan Sun, Junhang Zhang, Tianlong Ma, and Liang He.
\newblock A survey of human-in-the-loop for machine learning.
\newblock {\em Future Generation Computer Systems}, 2022.

\bibitem{gentile2019explore}
Anna~Lisa Gentile, Daniel Gruhl, Petar Ristoski, and Steve Welch.
\newblock Explore and exploit. dictionary expansion with human-in-the-loop.
\newblock In {\em The Semantic Web: 16th International Conference, ESWC 2019,
  Portoro{\v{z}}, Slovenia, June 2--6, 2019, Proceedings 16}, pages 131--145.
  Springer, 2019.

\bibitem{zhang2019invest}
Shanshan Zhang, Lihong He, Eduard Dragut, and Slobodan Vucetic.
\newblock How to invest my time: Lessons from human-in-the-loop entity
  extraction.
\newblock In {\em Proceedings of the 25th ACM SIGKDD International Conference
  on Knowledge Discovery \& Data Mining}, pages 2305--2313, 2019.

\bibitem{klie2020zero}
Jan-Christoph Klie, Richard~Eckart de~Castilho, and Iryna Gurevych.
\newblock From zero to hero: Human-in-the-loop entity linking in low resource
  domains.
\newblock In {\em Proceedings of the 58th annual meeting of the association for
  computational linguistics}, pages 6982--6993, 2020.

\bibitem{wallace2019trick}
Eric Wallace, Pedro Rodriguez, Shi Feng, Ikuya Yamada, and Jordan Boyd-Graber.
\newblock Trick me if you can: Human-in-the-loop generation of adversarial
  examples for question answering.
\newblock {\em Transactions of the Association for Computational Linguistics},
  7:387--401, 2019.

\bibitem{bartolo2020beat}
Max Bartolo, Alastair Roberts, Johannes Welbl, Sebastian Riedel, and Pontus
  Stenetorp.
\newblock Beat the ai: Investigating adversarial human annotation for reading
  comprehension.
\newblock {\em Transactions of the Association for Computational Linguistics},
  8:662--678, 2020.

\bibitem{bota2019semi}
Patr{\'\i}cia Bota, Joana Silva, Duarte Folgado, and Hugo Gamboa.
\newblock A semi-automatic annotation approach for human activity recognition.
\newblock {\em Sensors}, 19(3):501, 2019.

\bibitem{adaimi2019leveraging}
Rebecca Adaimi and Edison Thomaz.
\newblock Leveraging active learning and conditional mutual information to
  minimize data annotation in human activity recognition.
\newblock {\em Proceedings of the ACM on Interactive, Mobile, Wearable and
  Ubiquitous Technologies}, 3(3):1--23, 2019.

\bibitem{miu2015bootstrapping}
Tudor Miu, Paolo Missier, and Thomas Pl{\"o}tz.
\newblock Bootstrapping personalised human activity recognition models using
  online active learning.
\newblock In {\em 2015 IEEE International Conference on Computer and
  Information Technology; Ubiquitous Computing and Communications; Dependable,
  Autonomic and Secure Computing; Pervasive Intelligence and Computing}, pages
  1138--1147. IEEE, 2015.

\bibitem{sculley2007online}
D~Sculley.
\newblock Online active learning methods for fast label-efficient spam
  filtering.
\newblock In {\em CEAS}, volume~7, page 143, 2007.

\bibitem{abney2002bootstrapping}
Steven Abney.
\newblock Bootstrapping.
\newblock In {\em Proceedings of the 40th annual meeting of the Association for
  Computational Linguistics}, pages 360--367, 2002.

\bibitem{artstein2008inter}
Ron Artstein and Massimo Poesio.
\newblock Inter-coder agreement for computational linguistics.
\newblock {\em Computational linguistics}, 34(4):555--596, 2008.

\bibitem{scikit-learn-cohen}
scikit learn.
\newblock {\em Cohen’s kappa - scikit-learn}, 2022.
\newblock
  \url{https://scikit-learn.org/stable/modules/generated/sklearn.metrics.cohen_kappa_score.html},
  Last accessed on 2022-10-02.

\bibitem{brenner1999errors}
Steven~E Brenner.
\newblock Errors in genome annotation.
\newblock {\em Trends in Genetics}, 15(4):132--133, 1999.

\bibitem{hoelzemann2020digging}
Alexander Hoelzemann and Kristof Van~Laerhoven.
\newblock Digging deeper: towards a better understanding of transfer learning
  for human activity recognition.
\newblock In {\em Proceedings of the 2020 International Symposium on Wearable
  Computers}, pages 50--54, 2020.

\bibitem{reyes2016transition}
Jorge-L Reyes-Ortiz, Luca Oneto, Albert Sam{\`a}, Xavier Parra, and Davide
  Anguita.
\newblock Transition-aware human activity recognition using smartphones.
\newblock {\em Neurocomputing}, 171:754--767, 2016.

\bibitem{scholl2015wearables}
Philipp~M Scholl, Matthias Wille, and Kristof Van~Laerhoven.
\newblock Wearables in the wet lab: a laboratory system for capturing and
  guiding experiments.
\newblock In {\em Proceedings of the 2015 ACM International Joint Conference on
  Pervasive and Ubiquitous Computing}, pages 589--599, 2015.

\bibitem{stisen2015smart}
Allan Stisen, Henrik Blunck, Sourav Bhattacharya, Thor~Siiger Prentow,
  Mikkel~Baun Kj{\ae}rgaard, Anind Dey, Tobias Sonne, and Mads~M{\o}ller
  Jensen.
\newblock Smart devices are different: Assessing and mitigatingmobile sensing
  heterogeneities for activity recognition.
\newblock In {\em Proceedings of the 13th ACM conference on embedded networked
  sensor systems}, pages 127--140, 2015.

\bibitem{ioffe2015batch}
Sergey Ioffe and Christian Szegedy.
\newblock Batch normalization: Accelerating deep network training by reducing
  internal covariate shift.
\newblock In {\em International conference on machine learning}, pages
  448--456. PMLR, 2015.

\bibitem{bulling_2014}
Andreas Bulling, Ulf Blanke, and Bernt Schiele.
\newblock A tutorial on human activity recognition using body-worn inertial
  sensors.
\newblock {\em ACM Comput. Surv.}, 46(3), 2014.

\bibitem{ma2019bayesian}
Zhiheng Ma, Xing Wei, Xiaopeng Hong, and Yihong Gong.
\newblock Bayesian loss for crowd count estimation with point supervision.
\newblock In {\em Proceedings of the IEEE/CVF international conference on
  computer vision}, pages 6142--6151, 2019.

\bibitem{song2022learning}
Hwanjun Song, Minseok Kim, Dongmin Park, Yooju Shin, and Jae-Gil Lee.
\newblock Learning from noisy labels with deep neural networks: A survey.
\newblock {\em IEEE Transactions on Neural Networks and Learning Systems},
  2022.

\bibitem{smith2019comparing}
Mitchell~R Smith, Rifai Chai, Hung~T Nguyen, Samuele~M Marcora, and Aaron~J
  Coutts.
\newblock Comparing the effects of three cognitive tasks on indicators of
  mental fatigue.
\newblock {\em The Journal of psychology}, 153(8):759--783, 2019.

\bibitem{watkins2002implicit}
Philip~C Watkins.
\newblock Implicit memory bias in depression.
\newblock {\em Cognition \& Emotion}, 16(3):381--402, 2002.

\bibitem{chung2003exploring}
Janne Chung and Gary~S Monroe.
\newblock Exploring social desirability bias.
\newblock {\em Journal of Business Ethics}, 44:291--302, 2003.

\bibitem{aeronauticsNASATaskLoad1986}
NASA.
\newblock {NASA} task load index ({NASA}-{TLX}), version 1.0: {Paper} and
  pencil package.
\newblock 1986.

\bibitem{ghosh2015annotation}
Arindam Ghosh, Morena Danieli, and Giuseppe Riccardi.
\newblock Annotation and prediction of stress and workload from physiological
  and inertial signals.
\newblock In {\em 2015 37th annual international conference of the ieee
  engineering in medicine and biology society (embc)}, pages 1621--1624. IEEE,
  2015.

\bibitem{pal2020future}
Debajyoti Pal, Suree Funilkul, and Vajirasak Vanijja.
\newblock The future of smartwatches: assessing the end-users’ continuous
  usage using an extended expectation-confirmation model.
\newblock {\em Universal Access in the Information Society}, 19:261--281, 2020.

\bibitem{lieder2018anchoring}
Falk Lieder, Thomas~L Griffiths, Quentin~J M.~Huys, and Noah~D Goodman.
\newblock The anchoring bias reflects rational use of cognitive resources.
\newblock {\em Psychonomic bulletin \& review}, 25:322--349, 2018.

\bibitem{klayman1995varieties}
Joshua Klayman.
\newblock Varieties of confirmation bias.
\newblock {\em Psychology of learning and motivation}, 32:385--418, 1995.

\bibitem{min2016common}
Hyounae Min, Jeongdoo Park, and Hyun~Jeong Kim.
\newblock Common method bias in hospitality research: A critical review of
  literature and an empirical study.
\newblock {\em International Journal of Hospitality Management}, 56:126--135,
  2016.

\bibitem{wang2019attention}
Kun Wang, Jun He, and Lei Zhang.
\newblock Attention-based convolutional neural network for weakly labeled human
  activities’ recognition with wearable sensors.
\newblock {\em IEEE Sensors Journal}, 19(17):7598--7604, 2019.

\bibitem{wang2021sequential}
Kun Wang, Jun He, and Lei Zhang.
\newblock Sequential weakly labeled multiactivity localization and recognition
  on wearable sensors using recurrent attention networks.
\newblock {\em IEEE Transactions on Human-Machine Systems}, 51(4):355--364,
  2021.

\bibitem{yordanova2019challenges}
Kristina Yordanova.
\newblock Challenges providing ground truth for pervasive healthcare systems.
\newblock {\em IEEE Pervasive Computing}, 18(2):100--104, 2019.

\end{thebibliography}
\end{document}